\documentclass[journal]{IEEEtran}
\IEEEoverridecommandlockouts
\usepackage{cite}
\usepackage{amsmath,amssymb,amsfonts}
\usepackage{algorithmic}
\usepackage[linesnumbered,ruled,vlined]{algorithm2e}
\usepackage{graphicx}
\usepackage{textcomp}
\usepackage{xcolor}
\usepackage{subfigure}
\def\BibTeX{{\rm B\kern-.05em{\sc i\kern-.025em b}\kern-.08em
    T\kern-.1667em\lower.7ex\hbox{E}\kern-.125emX}}
\begin{document}

\title{Collaborative Optimization of Wireless Communication and Computing Resource Allocation based on Multi-Agent Federated Weighting Deep Reinforcement Learning}

\author{Junjie~Wu,
        Xuming~Fang,~\IEEEmembership{Senior Member,~IEEE},

\thanks{J. Wu and X. Fang are with Key Lab of Info Coding \& Transmission, Southwest Jiaotong University, Chengdu 610031, China. (E-mails: junjie\_wu@my.swjtu.edu.cn, xmfang@swjtu.edu.cn).

% G. Min is with the Department of Computer Science,University of Exeter, Exeter, EX4 4QF, U.K. (E-mail:g.min@exeter.ac.uk).

This work was supported in part by the NSFC under Grant 62071393, Sichuan Provincial Applied Basic Research Project under Grant 2020YJ0218, Fundamental Research Funds for the Central Universities under Grant 2682023ZTPY058.

In addition, this work has been submitted to the IEEE for possible publication. Copyright may be transferred without notice, after which this version may no longer be accessible.
}}

% \title{Multi-Agent Federated Weighted Deep Reinforcement Learning-based Computing and Communication Resource Synergistic Optimization
% % -oriented for Distributed Wireless System 
% % *\\
% % {\footnotesize \textsuperscript{*}Note: Sub-titles are not captured in Xplore and
% % should not be used}

% \author{\IEEEauthorblockN{1\textsuperscript{st} Junjie Wu}
% \IEEEauthorblockA{\textit{Information and Communication Engineering} \\
% \textit{Southwest Jiaotong University}\\
% Chengdu, China \\
% junjie\_wu@my.swjtu.edu.cn}
% \and
% \IEEEauthorblockN{2\textsuperscript{nd} Xuming Fang}
% \IEEEauthorblockA{\textit{Information and Communication Engineering} \\
% \textit{Southwest Jiaotong University}\\
% Chengdu, China \\
% xmfang@swjtu.edu.cn}
% }

\maketitle

\begin{abstract}
% In wireless communication scenarios, distributed deployment aligns better with actual application requirements. Therefore, in the development of intelligent wireless communication, we place greater emphasis on considering distributed learning methods. Existing studies have indicated that distributed learning only considers its own network state information, resulting in performance gaps compared to centralized training methods. To address the limitations, discrepancies, and privacy security concerns of distributed learning, we propose utilizing federated learning (FL) to optimize distributed network models in wireless systems. However, the traditional federated average training method fails to consider the heterogeneity of the distributed wireless system state. By adjusting the federated weighting coefficients based on heterogeneity, we can further optimize the performance of the federated network. The main focus of this paper is to utilize multi-agent federated deep reinforcement learning (MAFDRL) to tackle the joint optimization of computing and communication resources in next-generation wireless communication systems. By coordinating computing and communication strategies, we aim to address privacy security, system robustness, and performance optimization issues in distributed wireless systems.Extensive simulation experiments demonstrate that the proposed scheme outperforms baseline methods significantly in terms of performance improvement.

As artificial intelligence (AI)-enabled wireless communication systems continue their evolution, distributed learning has gained widespread attention for its ability to offer enhanced data privacy protection, improved resource utilization, and enhanced fault tolerance within wireless communication applications. Federated learning further enhances the ability of resource coordination and model generalization across nodes based on the above foundation, enabling the realization of an AI-driven communication and computing integrated wireless network. This paper proposes a novel wireless communication system to cater to a personalized service needs of both privacy-sensitive and privacy-insensitive users. We design the system based on based on multi-agent federated weighting deep reinforcement learning (MAFWDRL). The system, while fulfilling service requirements for users, facilitates real-time optimization of local communication resources allocation and concurrent decision-making concerning computing resources. Additionally, exploration noise is incorporated to enhance the exploration process of off-policy deep reinforcement learning (DRL) for wireless channels. Federated weighting (FedWgt) effectively compensates for heterogeneous differences in channel status between communication nodes. Extensive simulation experiments demonstrate that the proposed scheme outperforms baseline methods significantly in terms of throughput, calculation latency, and energy consumption improvement.

\end{abstract}

\begin{IEEEkeywords}
AI-enabled wireless communication systems, distributed learning, multi-agent federated weighting deep reinforcement learning (MAFWDRL), privacy protection
\end{IEEEkeywords}

\section{Introduction}
According to relevant studies \cite{b1,b2,b2_a,b3}, the current number of wireless communication access networks is significantly and continuously growing. A large number of wireless devices will be connected to these networks, resulting in massive data transmission and task computation requirements. The traditional cloud computing paradigm is not suitable for latency-sensitive applications and fails to meet the wireless communication latency-sensitive demands of future network systems. To address this issue, researchers have introduced mobile edge computing (MEC) into the field of wireless communication. Internet of things (IoT) devices can offload data and computational tasks to edge servers. This could have several advantages to the future wireless systems: 1) Processing computational tasks at edge devices or edge servers reduces round-trip data transmission time, which is crucial for latency-sensitive communication scenarios like autonomous driving, virtual reality (VR), and augmented reality (AR). 2) Edge computing nodes can locally process computational tasks, which reduce server load and time delay for data transmission from terminal devices to server. 3) Edge computing can decentralize computation loads, thereby lowering the probability of network congestion. 4) Deploying computational resources at the edge allows for local task computation, thereby reducing security risks associated with long distance transmitting privacy-sensitive data.

In addition, considering the distributed deployment characteristics of the future wireless network, if the centralized learning method is still adopted in the future network, this will cause the algorithm strategy to be unable to expand to large-scale network systems, and cannot cope with increasingly complex wireless network communication needs. Therefore, in order to meet the requirements of future distributed communication scenarios, research on network optimization for distributed communication problems is essential. In a distributed mobile communication scenario, effectively managing computing tasks and communication resources is a complex issue that needs to jointly consider computing task priorities, resource allocation, and scheduling strategies to achieve efficient computing and communication services. Tang \emph{et al.} \cite{b4} considered deployment of a deep learning model (deep learning, DL) to optimize the problem that is difficult to complete local training and upload weights to the edge server in time, which effectively reduces the computing delay. Zhao \emph{et al.} \cite{b5} introduced federated learning (FL) method into edge computing, and the performance impact of data distribution differences in FL was explored, and a wireless data sharing resource optimization method was also designed. For privacy-sensitive and privacy-insensitive user equipments (UEs) in distributed networks, Huang \emph{et al.} \cite{b6} adjusted an FL strategy to optimize computing resources allocation and reduce computing delay. Guo \emph{et al.} \cite{b7} considered the multi-agent deep reinforcement learning (MADRL) method in terms of communication optimization based on distributed learning, and the channel access problem is optimized based on the centralized training and distributed execution (centralized training with decentralized execution, CTDE) architecture.

However, when multiple nodes are involved in computing and communication tasks, it is necessary to consider the issue of coordinating data updates and synchronization among nodes to ensure the accuracy and consistency of calculations. In addition, task computing and communication will also affect each other. When the computing task delay is too long, the waiting time for communication will increase, and vice versa, which will directly degrade the overall performance of the system. In addition, privacy and security issues are also important and difficult challenges in wireless communication systems \cite{attachment_niyato}. Each node may face security risks such as tampering, eavesdropping, and unauthorized access during data transmission. For privacy-sensitive UEs with security risks, effective security measures must be taken to protect the confidentiality and integrity of the data.

Given the above problems, this paper proposes an optimization method of multi-agent federated deep reinforcement learning to solve the joint optimization problem of computing and communication resource allocation and simultaneously to ensure the privacy security of user communication. Our main contributions are as follows:
\begin{itemize}
\item We develop  an intelligent distributed computing and channel resource allocation method, using the MADRL method to train the local channel state, considering the dynamic optimization of computing and communication resources allocation of each node. The ultimate goal is to solve the joint non-convex optimization problem and coordinate computing and communication resource allocation to achieve optimal system performance.
\item	We propose a  new network framework MAFWDRL, which integrates FedWgt and MADRL. This framework takes into account state heterogeneity among training nodes and further improves distributed machine learning model performance. In addition, we design a  novel exploration noise function to fully explore the wireless channel status and further optimize the network model of the communication system.
\item	Considering privacy and security issues of wireless communication systems, we develop a variety of model training and selection methods for privacy-sensitive and privacy-insensitive UEs. Distributed computing and execution strategies are adopted when dealing with privacy-sensitive UEs, and privacy-sensitive UEs only consider performing computing tasks locally. UEs that are not sensitive to privacy can consider distributed or centralized computing and distributed execution strategies, and tasks can be sent back to the server for computing processing. This personalized service strategy can maximize the application efficiency of the intelligent model based on ensuring the needs of UEs.
\end{itemize}

The rest of the paper is organized as follows: Section II describes the details of the system model. Section III presents the proposed algorithm. Section IV reports and discusses the simulation results. Finally, Section V presents conclusions and future work.

\section{System Model and Problem Formulation}

We consider a typical wireless communication network serving different classes of devices, as shown in Fig. 1, including privacy-sensitive and privacy-insensitive UEs. Each wireless communication device works in a non-unique communication standard (to simulate the diversity of communication devices in reality). Both the UEs and BS of the communication system are equipped with a neural network (NN) to learn and optimize the communication and computing resource allocation. The UEs (actor networks) perform their own state collection and action determination, and the BS (actor and critic networks) perform task calculation, state collection, and evaluation of the actor value, to achieve the purpose of adaptively adjusting the operating parameters of the communication system.

\begin{figure}[]
  \begin{center}
    \scalebox{0.20}[0.20]{\includegraphics{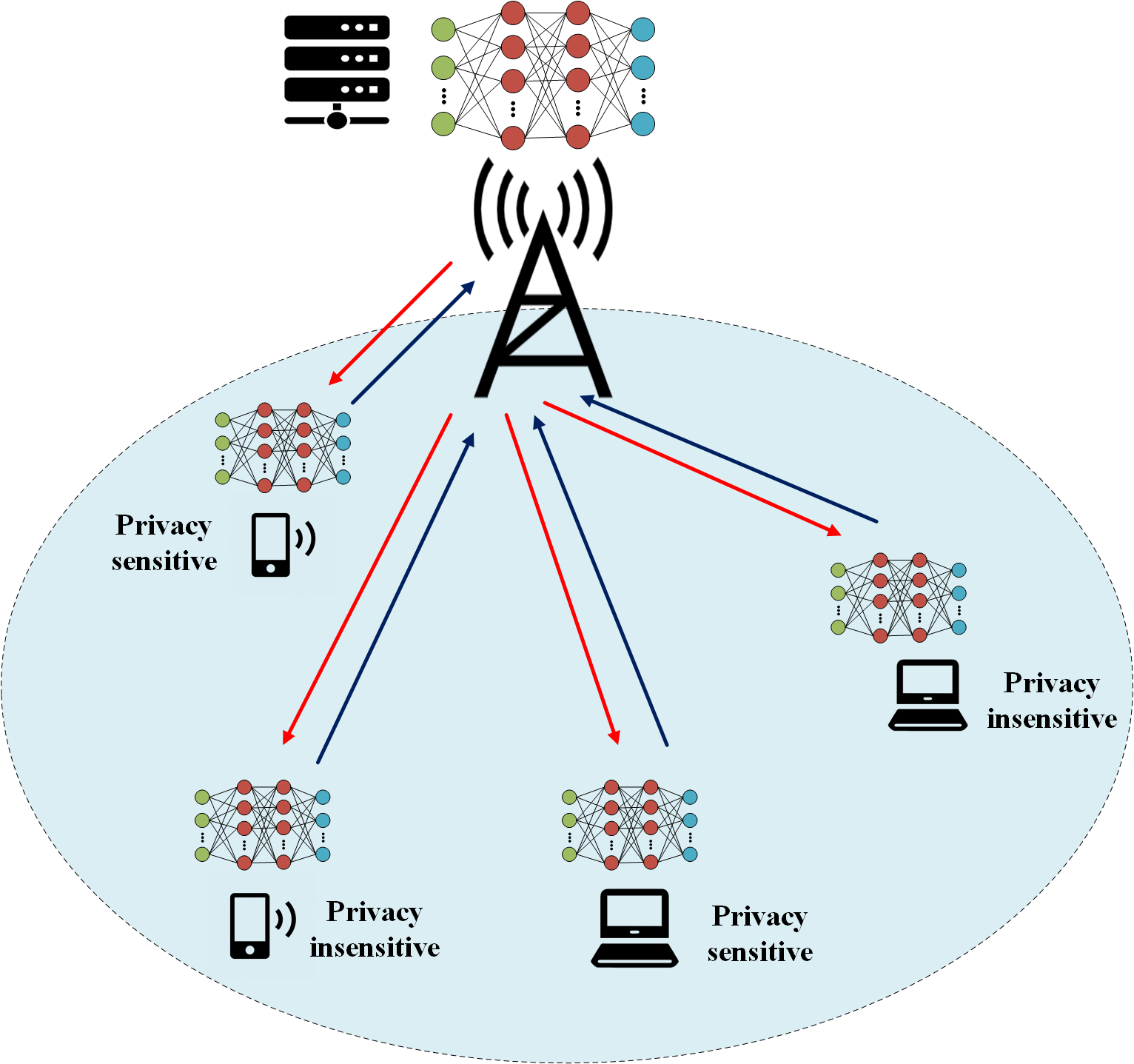}}
    \caption{Wireless communication scenario based on distributed learning.}
    \label{fig:1}
  \end{center}
\end{figure}

% \subsection{Mobility Model}
% We construct a mobility model of wireless communication based on NS3-simulator to simulate UEs dynamic access control. We divide the task period $T$ into $L$ time slots to observe a variation of wireless channel state information within the time slot length $\tau =T⁄L$. In addition, we randomly scatter several wireless access devices in a fixed area, including ${M}$ privacy-sensitive UEs and ${N}$ privacy-insensitive UEs. Wherein, the base station (BS) is located in the center of the service range to ensure that the communication coverage requirements of all UEs can be realized. The location information of the BS and the $k$-th UE at time slot $t_{i}$  is denoted as $\boldsymbol{s}=(x_{0},y_{0})$,$\boldsymbol{u}_{t_{i}}[k]=\left(x_{t_{i}}[k], y_{t_{i}}[k]\right)$. We consider that UE moves randomly with an angle of $\beta$ along the $X$-axis direction with a speed of $v[k]$, which are determined by user-provided random variables. Additionally, if UEs hit one of the communication boundaries, UEs will rebound on the boundary with a reflexive angle and speed. Then, the position of the $k$-th UE at the next moment $t_{i+1}$ is recorded as $\boldsymbol{u}_{t_{i+1}}[k]=\left(x_{t_{i}}[k]+v[k] \tau \cos \beta, y_{t_{i}}[k]+v[k] \tau \sin \beta\right)$. At this time, the Euclidean distance between the UE and the BS can be written as:

% \begin{equation}
% \boldsymbol{d}_{t_{i+1}}[\mathrm{k}]=\left\|\boldsymbol{u}_{t_{i+1}}[k]-\boldsymbol{s}\right\|_{2}, \forall i, k
% \end{equation}
% where $\| \boldsymbol{\cdot} \|_{2}$ represents the 2-norm.

\subsection{Channel Propagation Model}

To accurately simulate real and complex wireless communication scenarios, we consider the Log Distance Path Loss (LDPL) model \cite{b8} for our system. This model takes into account the random shadowing effects caused by obstacles such as hills, trees, and buildings, in different environments, to capture a broader range of propagation losses. Therefore, the path loss model between the BS and the $k$-th UE in the $i$-th time slot $t_{i}$ can be formulated as follows:
\begin{equation}
\left[P L\left(d_{0}\right)\right] d B=20 \log _{10}\left[\frac{4 \pi d_{0}}{\lambda}\right]
\end{equation}
% \begin{equation}
\begin{align}
\left[P L\left(\boldsymbol{d}_{t_{i}}[{k}]\right)\right] d B =& \left[P L\left(d_{0}\right)\right] d B+10 \gamma \log _{10}\left(\frac{\boldsymbol{d}_{t_{i}}[{k}]}{d_{0}}\right) \notag\\
&+\chi-\left[G_{t_{x}}(k)\right]d B i-\left[G_{r_{x}}(k)\right]d B i 
\end{align}
% \end{equation}
where $P L\left(d_{0}\right)$ called close-in reference distance, is obtained by using Friis path loss equation. The variable $\lambda$ exists in the path loss equation to account for the effective aperture of the receiving antenna, which is an indicator of the antenna’s ability to collect power. $d_{0}$ represents field measurements distance, that specifically denotes a value of 1m. $\gamma$ is the path loss exponent (PLE) that depends on the type of environment. $\chi $ is a zero-mean Gaussian distributed random variable with standard deviation $\sigma$ expressed in dB, used only when there is a shadowing effect. Then, the signal-to-noise ratio (SNR) received by the $k$-th UE can be expressed as follows:
\begin{equation}
\mathcal{S N R}_{k}[i]=\frac{P_{k}-P L\left(\boldsymbol{d}_{t_{i}}[\mathrm{k}]\right)}{\mathcal{N}_{0} b_{k}}
\end{equation}
where $P_{k}$ represents transmission power, the noise power spectral density is denoted as $\mathcal{N}_{0}$, and the $k$-th UE bandwidth is denoted as $b_{k}$. Consequently, the achievable rate of the distributed system during the $i$-th time slot can be expressed as:
\begin{equation}
\mathcal{R}_{k}[i]=b_{k} \log _{2}\left(1+\mathcal{S N R}_{k}[i]\right).
\end{equation}

The system takes into account the broadband transmission demands of the future wireless network and investigates the enhancements of MAFWDRL in terms of multi-parameter adjustment performance. Consequently, we consider SU-MIMO in the wireless communication system. Herein, the BS is equipped with $\lambda_{0}$ antennas, while the $k$-th UE possesses $\lambda_{k}$ antennas. During the $i$-th observation slot $\tau$, the channel transmission model of the $k$-th UE transmitting data can be represented as follows:
\begin{align}
D_{k}[i]=&\boldsymbol{C}\left[\begin{array}{c}
\mathrm{Y}_{1} \\
\mathrm{Y}_{2} \\
\vdots \\
\mathrm{Y}_{\lambda_{k}}
\end{array}\right]
\notag\\
=&\boldsymbol{C}\left(\left[\begin{array}{cccc}
h_{11} & h_{21} & \cdots & h_{\lambda_{0} 1} \\
h_{12} & h_{22} & \cdots & h_{\lambda_{0} 2} \\
\cdots & \cdots & \cdots & \cdots \\
h_{1 \lambda_{k}} & h_{2 \lambda_{k}} & \cdots & h_{\lambda_{0} \lambda_{k}}
\end{array}\right] \cdot\left[\begin{array}{c}
X_{1} \\
X_{2} \\
\vdots \\
X_{\lambda_{0}}
\end{array}\right]\right)
\end{align}
where $\mathrm{Y}_{\lambda_{k}}$ represents the received data of the $\lambda_{k}$-th spatial stream, while $\boldsymbol{C}=\underbrace{\left[\begin{array}{llll}
1 & 1 & \cdots & 1
\end{array}\right]}_{\lambda_{k}}$ denotes the vector employed to calculate the sum of $\left[\begin{array}{llll}
Y_{1} & Y_{2} & \cdots & Y_{\lambda_{K}}
\end{array}\right]^{T}$. The channel matrix associated with the $k$-th UE is symbolically represented as $\boldsymbol{H}_{k}[i]=\left[\begin{array}{cccc}
h_{11} & h_{21} & \cdots & h_{\lambda_{0} 1} \\
h_{12} & h_{22} & \cdots & h_{\lambda_{0} 2} \\
\cdots & \cdots & \cdots & \cdots \\
h_{1 \lambda_{k}} & h_{2 \lambda_{k}} & \cdots & h_{\lambda_{0} \lambda_{k}}
\end{array}\right]$. Furthermore, we consider that the amount of data in the transmission queue within the $i$-th time slot is denoted as $\boldsymbol{X}_{k}[i]=\left[\begin{array}{llll}
X_{1} & X_{2} & \cdots & X_{\lambda_{0}}
\end{array}\right]^{T}$, while the received data is denoted as $\boldsymbol{Y}_{k}[i]=\left[\begin{array}{llll}
Y_{1} & Y_{2} & \cdots & Y_{\lambda_{K}}
\end{array}\right]^{T}$. $D_{k}[i]$ represents the total data volume of the $k$-th user during the $i$-th time slot, the reality of throughput can be formally expressed as follows:  
\begin{equation}
    \mathcal{T}_{k}[i]=\frac{D_{k}[i]}{\tau}.
\end{equation}

Therefore, the actual throughput of overall system  $\mathcal{T}$ during the $T$ period can be expressed as:
\begin{equation}
    \mathcal{T}=\frac{D}{T}=\frac{\sum_{i=1}^{L} \sum_{k=1}^{M+N} D_{k}[i]}{T}.
\end{equation}

However, the inherent instability of wireless systems caused by factors such as system congestion, access contention, and channel state fluctuations, the actual system throughput and Shannon limit within the $i$-th time slot have the following constraints, which are denoted as:
\begin{equation}
    \mathcal{T}_{k}[i] \leq \mathcal{R}_{k}[i].
\end{equation}

\subsection{Computing Energy Consumption Model}
In the context of AI-enabled future wireless communication systems, it is crucial to establish an AI framework that covers user needs and can intelligently optimize communication and computing resource allocation. To accomplish this, we introduce a binary indicator known as $\delta_{k}=\{0,1\}$. Here, $\delta_{k}=0$ denotes the privacy-insensitive user category, while $\delta_{k}=1$ corresponds to the privacy-sensitive user category. It is imperative to address privacy and security concerns during the design phase of this model. Consequently, the computing architecture model, rooted in privacy and security considerations within this system, primarily focuses on adapting the following modules.

 \emph{\text { 1) } Privacy-insensitive computing model}: For $\delta_{k}=0$, the user type is categorized as privacy-insensitive. Consequently, as there are no privacy security concerns for this type of user, they have an option to select either a semi-centralized or decentralized training mode based on the current distributed learning paradigm. In this mode, the computing model takes into account the energy consumption associated with local actor NN training, server-side critic NN training energy consumption, task calculation energy consumption, as well as the influence of uploading and offloading training parameters on system performance. Therefore, we establish the initial CPU frequency of the $k$-th local UE as $f_{k}$, define the data size of each returned gradient vector (or environment observation space) in the UE as $\zeta_{k}$, and quantify the local computing task as $d_{k}$. In line with the network training process, we denote the number of sampling samples for the $i$-th round of batch gradient descent as $\varepsilon_{k}[i]$, the number of cycles required to process 1 bit of data as $r_{k}$, and the computed value of floating point operations per second (FLOPS) in a single cycle as $c$. Consequently, the energy consumption \cite{b6} of the actor NN local calculation energy consumption in the $i$-th time slot is expressed as:
\begin{equation}
 \bar{{P}_{k}^{A}}[i]={\kappa_{k} f_{k}^{3}}[i]
\end{equation}
\begin{equation}
 \bar{{T}_{k}^{A}}[i]=\frac{\left(\zeta_{k} \cdot {\varepsilon_{k}}[i]+d_{k}[i]\right) \cdot r_{k}}{f_{k}[i]}
\end{equation}
\begin{equation}
    \bar{{E}_{k}^{A}}[i]=\bar{{P}_{k}^{A}}[i] \bar{{T}_{k}^{A}}[i]=\kappa_{k}\left(\zeta_{k} \cdot {\varepsilon_{k}}[i]+d_{k}[i]\right) \cdot r_{k} \cdot {\kappa_{k} f_{k}^{2}}[i]
\end{equation}
where $\kappa_{k}$ is the CPU effective switched capacitance. In addition, the energy consumption \cite{b8_a} of the actor NN training model is typically denoted as:
\begin{equation}
    \bar{\bar{T}}_{k}^{A}[i]=\frac{f l o p s_{k}^{A}[i]}{f_{k}[i] c}
\end{equation}
\begin{equation}
    \bar{\bar{E}}_{k}^{A}[i]=\frac{\kappa_{k} f l o p s_{k}^{A}[i] f_{k}^{2}[i]}{c}
\end{equation}
 Furthermore, the system must also consider the energy consumption associated with computing by the critic NN on the server, which is specifically expressed as:
\begin{equation}
    \bar{{T}_{k}^{C}}[i]=\frac{\left(\rho_{k} \cdot \varepsilon_{k}[i]+d_{s}[i]\right) r_{s}}{f_{s}[i]}
\end{equation}
\begin{equation}
    \bar{E}_{k}^{C}[i]=\bar{P}_{k}^{C}[i] \bar{{T}_{k}^{C}}[i]=\kappa_{s} \cdot\left(\rho_{k} \cdot \varepsilon_{k}[i]+d_{s}[i]\right) r_{s} \cdot f_{s}^{2}[i]
\end{equation}
Similarly, the training energy overhead of the critic NN can be expressed as:
\begin{equation}
    \bar{\bar{T}}_{k}^{C}[i]=\frac{f l o p s_{s}^{C}[i]}{f_{s}[i] c}
\end{equation}
\begin{equation}
    \bar{\bar{E}}_{k}^{C}[i]=\frac{\kappa_{s} f l o p s_{s}^{C}[i] f_{s}^{2}[i]}{c}
\end{equation}
where the size of each channel state sample, representing the environmental observation space in $k$-th UE, is defined as $\rho_{k}$. We denote the number of sampling samples for the $i$-th round of batch gradient descent as $\varepsilon_{k}[i]$. Additionally, computational tasks are quantified as $d_{s}$, while the CPU frequency of the server is denoted as $f_{s}$. Furthermore, we represent the number of cycles required to process 1-bit data as $r_{s}$.

 \emph{\text { 2) } Privacy-sensitive computing model}: When the type $\delta_{k}=1$, the user type is categorized as privacy-sensitive. Privacy-sensitive users must carefully consider data security concerns, thus need to deploy NNs locally and the adoption of a decentralized distributed training architecture. This training architecture primarily addresses the computational energy consumption of the local NN network, as well as the performance implications of uploading and downloading NN models on the system. Then the energy consumption in the $i$-th time slot of NN local computing is expressed as:
\begin{equation}
    \tilde{P}_{k}^{A \& C}[i]=\kappa_{k} f_{k}^{3}[i]
\end{equation}
\begin{equation}
    \widetilde{T}_{k}^{A \& C}[i]=\frac{\left(\left(\zeta_{k}+\rho_{k}\right) \cdot \varepsilon_{k}[i]+d_{k}[i]\right) \cdot r_{k}}{f_{k}[i]}
\end{equation}
\begin{align}
    \tilde{E}_{k}^{A \& C}[i]=&\widetilde{P}_{k}^{A \& C}[i] \widetilde{T}_{k}^{A \& C}[i] \notag\\=&\kappa_{k}\left(\left(\zeta_{k}+\rho_{k}\right) \cdot \varepsilon_{k}[i]+d_{k}[i]\right) \cdot r_{k} \cdot f_{k}^{2}[i]
\end{align}
% where the CPU frequency of the $k$-th local UE with privacy-sensitive considerations is denoted as $f_{k}$.
The size of each return gradient in $k$-th agent, which belongs to an NN, is defined as $\zeta_{k}$. Furthermore, $\rho_{k}$ represents the size of the channel state in the wireless system. In terms of the network training process, $\varepsilon_k[i]$ represents the number of sampling samples utilized for batch gradient descent in the $i$-th round. Additionally, $r_{k}$ denotes the number of cycles required to process 1-bit. The local NN model training energy consumption is denoted as :
\begin{equation}
    \bar{\widetilde{T}}_{k}^{A \& C}[i]=\frac{f l o p s_{k}^{A}[i]+f l o p s_{k}^{C}[i]}{f_{k}[i] c}
\end{equation}
\begin{equation}
    \bar{\tilde{E}}_{k}^{A \& C}[i]=\frac{\kappa_{k}\left(f \operatorname{lops}_{k}^{A}[i]+f \operatorname{lops}_{k}^{C}[i]\right) f_{k}^{2}[i]}{c}.
\end{equation}
Clearly, based on the aforementioned computational model, the computation cost of the entire distributed system, which includes both privacy-sensitive and non-sensitive users, within the framework of the distributed learning architecture during the $i$-th time slot, can be expressed as: 
\begin{align}
    E[{i}]=&\sum_{k=1}^{{M}}\left(\bar{E}_{k}^{A}[i]+\bar{\bar{E}}_{k}^{A}[i]+\bar{E}_{k}^{C}[i] +\bar{\bar{E}}_{k}^{C}[i]\right) \notag\\+&\sum_{k=1}^{N}\left(\widetilde{E}_{k}^{A \& C}[i]+\bar{\widetilde{E}}_{k}^{A \& C}[i]\right)
\end{align}
where $E[{i}]$ refers to the energy consumption of the $i$-th time slot in the overall distributed wireless system. 
\subsection{Problem Formulation}
Without loss of generality, to realize the MAFWDRL application, this paper constructs a wireless communication environment following Wi-Fi standards using the NS3-simulator. It is worth mentioning that the actions and states of individual agents can be adjusted as necessary to suit Non-Wi-Fi standard wireless scenarios. The objective of this paper is to optimize communication and computing resource allocation in scenarios with varying degrees of communication sensitivity, aiming to minimize the dispersion of task weights by federated learning in the wireless system, minimize federated energy consumption or calculation latency, and maximize system throughput. The objective function is denoted as $\mathcal{O}$.  The communication network environment is influenced by factors such as aggregation frame length, contention window, and CPU frequency. In this environment, we define the decision variables as follows: $\mathcal{C W}=\left\{{\varepsilon}_{{k}}, \forall k\right\}$,$\mathcal{F} \mathcal{L}=\left\{{\xi}_{k}, \forall k\right\}$,$\boldsymbol{F}=\left\{{f}_{k},{f}_{s}, \forall k,\forall s\right\}$. The optimization problem of the system for the $i$-th training round is formulated as:
\begin{subequations}
\begin{equation}
   \max _{{f}, {\varepsilon}, {\xi}} \mathcal{O}=\frac {\mathcal{T}[i]}{{E}[i]} \notag
\end{equation}
\begin{equation}
s.t.  0 \leqslant f_{k} \leqslant f_{k}^{\max }, k=1, \ldots, M
\end{equation}
\begin{equation}
0 \leqslant f_{s} \leqslant f_{s}^{\max }
\end{equation}
\begin{equation}
    0 \leqslant \varepsilon_{k} \leqslant \varepsilon_{k}^{\max }, k=1, \ldots, M
\end{equation}
\begin{equation}
0 \leqslant \xi_{k} \leqslant \xi_{k}^{\max }, k=1, \ldots, M
\end{equation}
\begin{equation}
    0 \leqslant \sum \xi_{k} \leqslant D_{k}^{\max }, k=1, \ldots, M
\end{equation}
\begin{equation}
    0 \leqslant T_{k \text {,insen }} \leqslant T_{k \text {,insen }}^{\text {max }}, k=1, \ldots, M
\end{equation}
\begin{equation}
    0 \leqslant T_{k, s e n} \leqslant T_{k, s e n}^{\max }, k=1, \ldots, M
\end{equation}
\end{subequations}
where $\mathcal{T}[i]$ denotes system throughput in $i$-th time slot, and  calculation energy consumption is denoted as $E[i]$. Constraints (24a) and (24b) define the limits on CPU frequency for STAs and AP, while constraints (24c) and (24d) specify the limits on CW and aggregate frame length in the action space for channel status. Additionally, constraint (24e) presents the maximum queue length limit for channel transmission. Constraints (24f) and (24g), respectively, guarantee the maximum computation delay for privacy-insensitive and privacy-sensitive tasks, thereby ensuring timely system responsiveness and preventing task failures or performance degradation caused by exceeding the expected computation time.

We consider that the wireless system is dynamic and intricate, leading to unpredictable communication conditions. To guarantee the robustness of the system, it is crucial to ensure its adaptability. Evidently, the interdependence among the variables $\mathcal{C W}$, $\mathcal{F L}$, and $\boldsymbol{F}$ renders the objective function $\mathcal{O}$ as a non-convex optimization problem for the wireless system. Consequently, we employ the MAFWDRL method to acquire the optimal action strategy for the system, thereby enabling the objective function $\mathcal{O}$ to obtain an optimal action at any given point in time.
\section{resource allocation and energy consumption optimization based on MAFWDRL }

In this section, we present the MAFWDRL algorithm, which combines FedWgt and MADRL to address the non-convex optimization problem in wireless systems. The algorithm is intelligently designed for wireless systems, as illustrated in Fig. 2. Furthermore, we examine the impact of privacy issues on the algorithm architecture, discuss the advantage of the FedWgt strategy, and design the explore noise function for MADRL.

\begin{figure*}[]
  \begin{center}
    \scalebox{0.3}[0.25]{\includegraphics{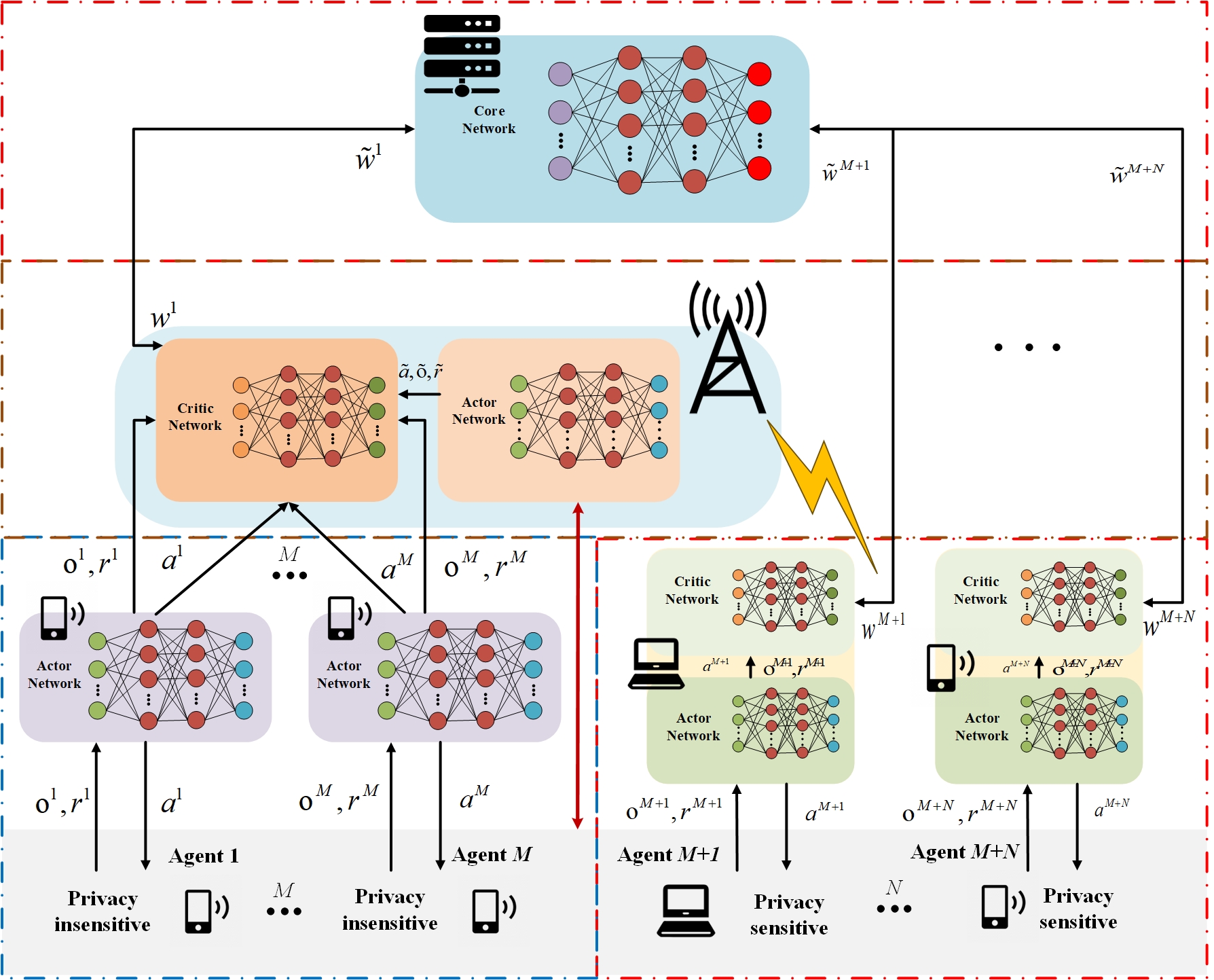}}
    \caption{Resource allocation optimization architecture based on MAFWDRL for distributed wireless networks.}
    \label{fig:4}
  \end{center}
\end{figure*}

\subsection{Federated Weighting Design}
The discrepancy in channel states between different UEs is observed in wireless communication systems, which are influenced by multiple factors, including noise, interference, communication distance, and resource allocation. The performance of FL is affected by the data state distribution (IID and Non-IID)  \cite{b9,b10,b10_a}. Based on this, we propose to assess the quality of channels by defining an abstract reward for the system, $R=\boldsymbol{CHX}$.

We consider the communication states between each training round to be mutually independent. The channel state for the $k$-th agent during the training process is denoted as $\boldsymbol{H}_{k}$, and the probability of the state occurring is denoted as $P\left\{X=\boldsymbol{H}_{{k}}\right\}$. The local reward of $k$-th agent in the wireless system is $R_{k}=\boldsymbol{CH_{k}X_{k}}$. The state of the global channel characteristic is denoted as $\boldsymbol{H}_{g}$, and the probability of the state occurring is $P\left\{Y=\boldsymbol{H}_{{g}}\right\}$. Consequently, the global state reward in the wireless system is $R_{g}=\boldsymbol{CH_{g}X_{g}}$. Therefore, the distributed learning expected reward for the $k$-th agent under the current channel state can be expressed as:
% \begin{equation}
%     \mathbb{E}\left\{R_{k}\right\}=R_{k} P\left\{X=\boldsymbol{H}_{\boldsymbol{k}}\right\}=\boldsymbol{C H}_{\boldsymbol{k}} \boldsymbol{X}_{\boldsymbol{k}} P\left\{X=\boldsymbol{H}_{\boldsymbol{k}}\right\}
% \end{equation}
\begin{align}
    \mathbb{E}\left\{R_{k}\right\} &=R_{k} P\left\{X=\boldsymbol{H}_{\boldsymbol{k}}\right\}
    \notag\\
    &=\boldsymbol{C H}_{\boldsymbol{k}} \boldsymbol{X}_{\boldsymbol{k}} P\left\{X=\boldsymbol{H}_{\boldsymbol{k}}\right\}
\end{align}
where $\boldsymbol{X}_{k}$ is the amount of data to be transmitted. Similarly, in this case, if centralized learning is adopted, the expected reward that should be pursued by the $k$-th agent is denoted as:
% \begin{equation}
%     \mathbb{E}\left\{R_{g}\right\}=\frac{1}{M+N} R_{g} P\left\{Y=\boldsymbol{H}_{\boldsymbol{g}}\right\}=\frac{1}{M+N} \boldsymbol{C H}_{\boldsymbol{g}} \boldsymbol{X}_{\boldsymbol{g}} P\left\{Y=\boldsymbol{H}_{\boldsymbol{g}}\right\}
% \end{equation}
\begin{align}
     \mathbb{E}\left\{R_{g}\right\} &=\frac{1}{M+N} R_{g} P\left\{Y=\boldsymbol{H}_{\boldsymbol{g}}\right\}
     \notag\\
     &=\frac{1}{M+N} \boldsymbol{C H}_{\boldsymbol{g}} \boldsymbol{X}_{\boldsymbol{g}} P\left\{Y=\boldsymbol{H}_{\boldsymbol{g}}\right\}
\end{align}
where $\boldsymbol{X}_{g}$ denotes the aggregate data volume intended for global transmission, and $M+N$ represents the total number of UEs. The deviation between local and global states is likely to result in a decrease in model accuracy as a consequence of NN model parameter $\theta$. Within each training round, the impact of this factor on the accuracy of the neural network model can be expressed as follows:
\begin{equation}
    G(\theta, \boldsymbol{H} \boldsymbol{X})=\mathbb{E}_{\boldsymbol{H} \sim P(X)}\{\nabla g(\theta, \boldsymbol{H} \boldsymbol{X})\}
\end{equation}
% \begin{equation}
%     \psi=\left\|G\left(\theta, \boldsymbol{H}_{\boldsymbol{k}} \boldsymbol{X}_{\boldsymbol{k}}\right)-G\left(\theta, \boldsymbol{H}_{g} \boldsymbol{X}_{g}\right)\right\|_{1} \leq \sum_{k=1}^{M}\left\|\mathbb{E}\left\{R_{k}\right\}-\mathbb{E}\left\{R_{g}\right\}\right\|_{1}
% \end{equation}
\begin{align}
    \psi &=\left\|G\left(\theta, \boldsymbol{H}_{\boldsymbol{k}} \boldsymbol{X}_{\boldsymbol{k}}\right)-G\left(\theta, \boldsymbol{H}_{g} \boldsymbol{X}_{g}\right)\right\|_{1} 
    \notag\\
    &\leq \sum_{k=1}^{M}\left\|\mathbb{E}\left\{R_{k}\right\}-\mathbb{E}\left\{R_{g}\right\}\right\|_{1}
\end{align}
where $G(\theta, \boldsymbol{H} \boldsymbol{X})$ represents the model gradient of the current training round, and $\| \boldsymbol{\cdot} \|_{1}$ is 1-norm. Based on the corollary in \cite{b5}, the difference between the local and global system states can be captured by establishing the correlation between federation weight and loss accuracy. Consequently, based on reference [6, Corollary 1], the model accuracy loss caused by the discrepancy of channel states can be refined as follows:
\begin{equation}
    \mathcal{Z}(\boldsymbol{H} \boldsymbol{X})=\tau \sum_{k=1}^{M} w_{k} \psi_{k}
\end{equation}
where $\tau=\frac{\eta\left[1-\left(\sqrt{1+\eta^{2} \beta^{2}-2 \eta \lambda}\right)^{\varepsilon}\right]}{1-\sqrt{1+\eta^{2} \beta^{2}-2 \eta \lambda}}$, $\eta$ is the learning rate, $\beta$ and $\lambda$ are the derivable constraint coefficients, representing the Lipschitz and smooth assumptions respectively. Modifying the weight parameter $w_{k}$ can alleviate the discrepancy between the local and global states, which can reduce the upper bound of the model accuracy loss $\mathcal{Z}$. Hence, to mitigate the impact of this discrepancy, we define the weight of FedWgt as follows [6, eq.(17)]:
\begin{equation}
    w_{k}=\frac{1 / \psi_{k}^{2}}{\sum_{n=1}^{M}\left(1 / \psi_{n}^{2}\right)}
\end{equation}
where $w_{k}$ is the FedWgt coefficient.
\subsection{Exploration Noise Design for Off-policy DRL}
Introducing exploration noise is an effective approach to address the limited exploration capability in off-policy DRL. During the learning process of NN, our objective is to encourage extensive exploration of the environment by the early-stage network, enabling the collection of a diverse range of environmental states. With the exploration progressing, the scale of noise decreases at an accelerated rate. As the later-stage NN network approaches convergence, the noise level tends to a lower bound for the exploration. Thus, it is essential to design a noise function $f(x)$ that ensures the rate of change increases gradually. This strategy facilitates reasonable exploration and aligns better with the NN learning characteristics. As a result, several conditions need to be met in the design of the noise function $f(x)$.

1) For a function defined on set $B$, if for any points $x_{0}$, $x_{1}$, and $x_{2}$ on the function where $x_{0}<x_{1}<x_{2} $ hold, then the condition 
% \begin{equation}
%     \left(\frac{\partial f\left(x_{0}\right)}{\partial x_{0}}-\frac{\partial f\left(x_{1}\right)}{\partial x_{1}}\right) /\left|x_{0}-x_{1}\right|<\left(\frac{\partial f\left(x_{1}\right)}{\partial x_{1}}-\frac{\partial f\left(x_{2}\right)}{\partial x_{2}}\right) /\left|x_{1}-x_{2}\right|
% \end{equation}
\begin{equation}
    \frac{\left(\frac{\partial f\left(x_{0}\right)}{\partial x_{0}}-\frac{\partial f\left(x_{1}\right)}{\partial x_{1}}\right)}{\left|x_{0}-x_{1}\right|}<\frac{\left(\frac{\partial f\left(x_{1}\right)}{\partial x_{1}}-\frac{\partial f\left(x_{2}\right)}{\partial x_{2}}\right)}{\left|x_{1}-x_{2}\right|}
\end{equation}
is always satisfied.

2) The noise decay curve defined on set $B$ is monotonically decreasing and the rate of decrease is monotonically increasing. Within the domain of $B$, it exhibits a concave function with gradually increasing and monotonically decreasing rate of change, expressed as:
\begin{equation}
    \frac{\partial^{2} f(x)}{\partial^{2} x}<{0}  \text { and } \frac{\partial f(x)}{\partial x}<{0}.
\end{equation}
% where $\partial^{2}$ is the 2-derivative, $\frac{\partial^{2} f(x)}{\partial^{2} x}<{0}$ denotes that $f(x)$ is a concave function, and $\frac{\partial f(x)}{\partial x}<{0}$ represents that $f(x)$ is monotonically decreasing.

\subsection{MAFWDRL algorithm design for distributed wireless system}
From a mathematical perspective, a multi-agent task can be defined as a decentralized partially observable Markov decision process (Dec-POMDP) \cite{b11,b12,b13}, where each agent independently selects actions to react to the environment ultimately. The Markov game of $N+M$ agents can be defined as the set of states $\mathcal{S}$, including the set of action spaces $\mathcal{A}=\left\{a_{1}, a_{2}, \cdots, a_{N+M}\right\}$, and the set of observation spaces $O =\left\{o_{1}, o_{2}, \cdots, o_{N+M}\right\} $, where each agent selects actions based on the policy $\pi\left(a_{k} \mid o_{k} ; \theta_{k}\right) \rightarrow a_{k}$. In addition, this paper considers the issue of local training discrepancies between $o_{1}, o_{2}, \cdots, o_{N+M}$, and $\mathcal{S}$, and introduces FedWgt $\mathcal{W}=\left\{w_{1}, w_{2}, \cdots, w_{N+M}\right\}$ to integrate the distributed NN models. Thus, the multi-agent policy can be adjusted as follows: $\pi\left(\tilde{a}_{k} \mid o_{k} ; w_{k}, \theta_{k}\right) \rightarrow \tilde{a}_{k}$, where $\tilde{\mathcal{A}}=\left\{\tilde{a}_{1}, \tilde{a}_{2}, \cdots, \tilde{a}_{N+M}\right\}$. The state transition equation $\mathcal{E}$ for the entire environment can be represented as: $\mathcal{S} \times \tilde{\mathcal{A}} \rightarrow \mathcal{S}$. At this point, the mapping between actions and states of the $i$-th agent under the current conditions can be directly characterized by rewards: $o_{k} \times \tilde{a}_{k} \rightarrow r_{k}$, where $\mathcal{R} = \left\{r_{1}, r_{2}, \cdots, r_{N+M}\right\}$. The goal of each agent is to maximize long-term expected returns, i.e., $\mathcal{R}_{k}=\sum_{t=0}^{T} \gamma^{t} r_{k}^{t}$, where $\gamma$ is the discount factor, and $T$ is a complete wireless task period in the wireless system. The Dec-POMDP problem in this paper, specifically in the context of distributed wireless systems, can be represented as a tuple $\mathcal{G}=(\mathcal{S}, \tilde{\mathcal{A}}, \mathcal{R}, \mathcal{E}, O, \mathcal{W}, \gamma)$.

 \emph{\text { 1) } Action Space $\mathcal{A}$}:
 In this paper, the optimization of MAC and CPU parameters to optimize the access and computation efficiency of the wireless channel is considered. The optimization is performed using an NN model, which takes into account the learned channel feature $\boldsymbol{H}$ and the state of the task volume $\boldsymbol{X}$. As a result, the action of agent $k\in\left\{1,2,\cdots,N+M\right\}$ at slot $t$ is defined as $a_{k}^{t} \in \mathcal{A} \triangleq\left[\mathcal{C W}, \mathcal{F} \mathcal{L}, \mathcal{F}_{s}, \mathcal{F}_{c}\right]$, where $\mathcal{C W}$ represents the contention window, and $\mathcal{F} \mathcal{L}$ refers to the aggregation of frame length. Additionally, $\mathcal{F}_{s}$ and $\mathcal{F}_{c}$ denote the CPU computing frequency of the server and clients, respectively.
 
 \emph{\text { 2)} Observation Space $\mathcal{S}$}:
 The observation space in the time slot $t$ is focused on by each agent $k\in\left\{1,2,\cdots,N+M\right\}$ in order to capture the changing environmental states within the system. This observation space enables the direct reflection of the communication state information of the wireless system. The information is primarily composed of the following three parts: 1) SNR of each agent is considered (ignore the interference). The SNR is influenced by factors such as user movement and environmental noise, thereby allowing for the intuitive depiction of the system's communication quality. 2) The packet loss rate of the system is examined. With the number of ACKs received by the device associated with agent $k$ ($i.e.$ $ c_{k}$) and the number of packets sent by agent $k$ ($i.e.$ $ d_{k}$) within the time slot $t$, the packet loss rate can be calculated as $\mathcal{P}_{k}=\frac{d_{k}-c_{k}}{d_{k}}$. This indicator provides insights into the extent of data loss during transmission. 3) The percentage of idle time is taken into account. The proportion of data transmission time during the TXOP period by agent $k$ ($i.e.$ $ t_{k}$), denoted as $v_{k}$, is monitored. Utilizing the formula $\mathcal{L}_{k}=\frac{t_{i-} v_{k}}{t_{k}}$ allows for the estimation of the proportion of idle time during transmission for agent $k$. Then, the local state of agent $k$ at slot $t$ is denoted as:
 \begin{equation}
     o_{k}^{t}=\left[\mathcal{S} \mathcal{N} \mathcal{R}_{\boldsymbol{k}}^{t}, \mathcal{P}_{\boldsymbol{k}}^{\boldsymbol{t}}, \mathcal{L}_{\boldsymbol{k}}^{\boldsymbol{t}}\right].
 \end{equation}
 
At the present moment, the observation space of all agents is represented as $O=\left[o_{1}^{t}, o_{2}^{t}, \cdots, o_{N+M}^{t}\right]$. Furthermore, the global state information, in the context of the off-policy, can be denoted as $\mathcal{S}= \left[\mathcal{A}, O\right]$, where $\mathcal{A}$ represents the action space of all agents.

 \emph{\text { 3)} Reward $\mathcal{R}$}:
  The optimization objective function is defined as $\mathcal{O}=\frac{\mathcal{T}}{E}$, where $\mathcal{O}$ is strongly related to the reward, serving as a positive indicator for evaluating network performance. Furthermore, when the system exceeds the defined maximum permissible computing latency threshold, a penalty must be imposed against the DRL model. This penalty is realized by adjusting the reward to a negative value (\emph{i.e.}, $r=-T_{k}$). Such an adjustment aims to instruct the DRL model in recognizing and quantifying the adverse effects that incorrect actions can have on the system's performance. Consequently, it is inferred that a larger delay signifies a greater deviation of the network from the correct direction. When prioritizing energy consumption, the system demands lower energy consumption within the computing delay limit to achieve an improved overall system reward. In addition, in order to eliminate the impact of numerical deviations of $T$, $\mathcal{T}$ and $E$ on the DRL model as much as possible, we use the arctangent normalization function (\emph{i.e.}, $z(x)=\frac{2}{\pi} \arctan (x)$) to normalize the reward $r$. Given these considerations, the rewards for wireless systems are defined as follows: 
  \begin{equation}
      r_{k}=\left\{\begin{array}{ll}
z(\frac{\mathcal{T}_{k}^{\delta_{k}=0}}{{E}_{k}^{\delta_{k}=0}}) , & \text { if } T_{k}^{\delta_{k}=0} \leq T_{k, i n s e n}^{\max }  \\
z(-T_{k}^{\delta_{k}=0}), & \text { else }  \text { if }  T_{k}^{\delta_{k}=0}>T_{k,  i n s e n }^{\max } \\
z(\frac{\mathcal{T}_{k}^{\delta_{k}=1}}{{E}_{k}^{\delta_{k}=1}}), & \text { else }  \text { if }  T_{k}^{\delta_{k}=1} \leq T_{k, s e n}^{\max } \\
z(-T_{k}^{\delta_{k}=1}), & \text { otherwise }
\end{array}\right.
  \end{equation}
where $\mathcal{T}_{k}$, $T_{k}$, and ${E}_{k}$  represent the throughput, computing latency, and computing energy consumption of the $k$-th STA, respectively. 

The objective of our distributed wireless system is to optimize the throughput and calculation latency or calculation energy consumption of the overall system. If the actions are entirely independent, certain constraints will arise that impact the system as a whole. Taking these considerations into account, we propose a system optimization framework based on the MAFWDRL method for wireless systems with privacy-sensitive and privacy-insensitive requirements. We consider soft MADRL under the off-policy mechanism as the core optimization algorithm. Our algorithm, as shown in Algorithm 1, consists of three main components: 1) Privacy-sensitive users learn actor NN and critic NN based solely on local information. 2) Privacy-insensitive users, having no concerns about privacy issues, would benefit more from prioritizing the semi-centralized training mode to enhance system performance. 3) A FedWgt learning approach is employed to optimize the global NN model.

 \emph{\text { 1) } Fully decentralized MADRL algorithm for privacy-sensitive}:
 Due to data security constraints and the privacy concerns of sensitive users (represented by the set $\boldsymbol{J} \in\left\{j_{1}, j_{2}, \cdots, j_{{M}}\right\}$, where the identifier $\delta = 1$), a fully decentralized soft  MADRL optimization algorithm is considered more suitable for communication applications. In this context, the Critic network parameters of ${M}$ agents are denoted as $\boldsymbol{\theta}^{c}=\left\{\vartheta_{j_{1}}^{c}, \vartheta_{j_{2}}^{c}, \cdots, \vartheta_{j_{{M}}}^{c}\right\}$, and the policy network parameter as $\boldsymbol{\theta}^{p}=\left\{\vartheta_{j_{1}}^{p}, \vartheta_{j_{2}}^{p}, \cdots, \vartheta_{j_{{M}}}^{p}\right\}$. The action is governed by the policy function $\boldsymbol{u}\left(\boldsymbol{o} \mid \boldsymbol{\theta}^{p}\right)=\left\{u_{j_{1}}\left(o_{j_{1}} \mid \vartheta_{j_{1}}^{p}\right), u_{j_{2}}\left(o_{j_{2}} \mid \vartheta_{j_{2}}^{p}\right), \cdots, u_{j_{{M}}}\left(o_{j_{{M}}} \mid \vartheta_{j_{{M}}}^{p}\right)\right\}$. The updated formula for the training of each critic is as follows:
    \begin{equation}
       y_{j_{k}}^{i}=r_{j_{k}}^{i}+\gamma \widetilde{Q}_{j_{k}}^{i+1}\left(o_{j_{k}}^{i+1}, \tilde{u}_{j_{k}}^{i+1}\left(o_{j_{k}}^{i+1} \mid \widetilde{\vartheta}_{j_{k}}^{{p}}\right) \mid \widetilde{\vartheta}_{j_{k}}^{c}\right)
    \end{equation}
    \begin{align}
        \mathcal{L}=\frac{1}{\mathbb{S}} \sum_{i}\left(y_{j_{k}}^{i}-Q_{j_{k}}^{i}\left(o_{j_{k}}^{i}, a_{j_{k}}^{i} \mid \vartheta_{j_{k}}^{c}\right)\right)^{2}
    \end{align}
where $i$ represents the $i$-th time slot, $\mathbb{S}$ represents mini-batch of samples, and the target network parameter denoted as $\widetilde{\vartheta}$, is subject to updates according to the following equation: 
\begin{equation}
    \widetilde{\vartheta}=\varphi \vartheta+(1-\varphi) \widetilde{\vartheta}
\end{equation}
 where $\varphi$ represents the learning coefficient. Similarly, the updated formula based on Q-value of agent $j_{k}$ for the training of each actor is as follows:
 \begin{align}
     \nabla_{\vartheta^{p}_{j_{k}}} \mathcal{J} \approx \frac{1}{\mathbb{S}} \sum_{i} &\nabla_{a_{j_{k}}} Q_{j_{k}}^{i}\left(o_{j_{k}}^{i}, a_{j_{k}}^{i} \mid \vartheta^{c}_{j_{k}}\right) \mid_{o_{j_{k}}^{i}, a_{j_{k}}^{i}=\mu\left(o_{j_{k}}^{i}\right)} \notag
     \\
     &\times \nabla_{\vartheta^{p}_{j_{k}}} \mu\left(o_{j_{k}}^{i} \mid \vartheta_{j_{k}}^{p}\right)\mid_{o_{j_{k}}^{i}}
  \end{align}
 where $\nabla_{\vartheta^{p}_{j_{k}}} \mathcal{J}$ is the policy gradient of the NN model for sensitive users.
 
 \emph{\text { 2) } Semi-centralized MADRL algorithm for privacy-insensitive}:
    For ${N}$ privacy-insensitive users $\boldsymbol{L} \in\left\{l_{1}, l_{2}, \cdots, l_{{N}}\right\}$, with the identifier $\delta=0$, regardless of data security constraints, sufficient state information can be provided for Critic training. Thus, a model framework featuring centralized training and distributed execution is more conducive to the learning of the system environment by the privacy-insensitive users (i.e., semi-centralized MADRL). The observation space of the $k$-th privacy-insensitive agent, denoted as $l_{k}$, is represented as $\boldsymbol{O}_{l_{k}}=\left(o_{l_{1}}, o_{l_{2}}, \cdots, o_{l_{{N}}}\right)$. The action of each agent is denoted as $a_{l_{1}}, a_{l_{2}}, \cdots, a_{l_{{N}}}$, and the policy function follows $\boldsymbol{u}\left( \boldsymbol{o} \mid \boldsymbol{\theta}^{p}\right)=\left\{u_{l_{1}}\left(o_{l_{1}} \mid \vartheta_{l_{1}}^{p}\right), u_{l_{2}}\left(o_{l_{2}} \mid \vartheta_{l_{2}}^{p}\right), \cdots, u_{l_{{N}}}\left(o_{l_{{N}}} \mid \vartheta_{l_{{N}}}^{p}\right)\right\}$. Therefore, the centralized action-value function can be expressed as $\left.Q_{l_{k}}\left(\boldsymbol{O}_{l_{k}} ; a_{l_{1}}, a_{l_{2}}, \cdots, a_{l_{N}}\mid \vartheta_{l_{k}}^{c}\right)\right|_{a_{l_{k}}=u\left(o_{l_{c}} \mid \vartheta_{l_{c}}^{p}\right)}$.The updated formula for the training of centralized critic NN is as follows:
    \begin{equation}  y_{l_{k}}^{i}=r_{l_{k}}^{i}+\left.\gamma \bar{Q}_{l_{k}}^{i+1}\left(\boldsymbol{O}_{l_{k}}^{i+1}, \bar{a}_{l_{1}}, \bar{a}_{l_{2}}, \ldots, \bar{a}_{l_{{N}}} \mid \bar{\vartheta}_{l_{k}}^{c}\right)\right|_{\bar{a}_{l_{c}}=\bar{u}_{l_{c}}\left(o_{j_{c}} \mid \bar{\vartheta}_{l_{c}}^{p}\right)}
    \end{equation}
    \begin{align}   &\mathcal{L}\left(\vartheta_{l_{k}}\right)= \notag \\
    &\frac{1}{\mathbb{S}} \sum_{i}\left(y_{l_{k}}^{i}-\left.Q_{l_{k}}^{i}\left(\boldsymbol{O}_{l_{k}}^{i}, a_{l_{1}}, a_{l_{2}}, \cdots, a_{l_{\mathrm{N}}} \mid \vartheta_{l_{k}}^{c}\right)\right|_{a_{l_{c}}=u_{l_{c}}\left(o_{l_{c}} \mid \vartheta_{l_{c}}^{p}\right)}\right)^{2}
    \end{align}
    % where the target network parameters, denoted as $\bar{\vartheta}$, while the Formula (38) ensures that the parameter updates adhere to the prescribed criteria.
    The training update formula, within the decentralized architecture, can be expressed as follows:
    \begin{align}
   &\nabla_{\theta_{l_{k}}} \mathcal{K} \approx \frac{1}{\mathbb{S}} \sum_{i} \nabla_{\vartheta_{l_{k}}^{p}} \mu_{l_{k}}\left(o_{l_{k}}^{i} \mid \vartheta_{l_{k}}^{p}\right) 
   \notag \\
   &\times \nabla_{a_{l_{k}}} Q_{l_{k}}^{i}\left(\boldsymbol{O}_{l_{k}}^{i}, a_{1}^{i}, \ldots, a_{l_{k}}, \ldots, a_{l_{\mathrm{N}}}^{i} \mid \vartheta_{l_{k}}^{c}\right)\mid_{a_{l_{c}}=u_{l_{c}}\left(o_{l_{c}} \mid \vartheta_{l_{c}}^{p}\right)}
    \end{align}
    where $\nabla_{\theta_{l_{k}}} \mathcal{K}$ is the policy gradient of the NN model for insensitive users.

 \emph{\text { 3) } Model Ensemble Based on FedWgt}:
    The fully centralized DRL for privacy-sensitive users and the semi-centralized DRL weighted integration strategy for privacy-insensitive users can be expressed as follows:
    \begin{equation}
\Theta_{k}=\sum_{k=1}^{{M}} w_{j_{k}} \vartheta_{j_{k}}+\sum_{k=1}^{{N}} w_{l_{k}} \vartheta_{l_{k}}
    \end{equation}

\begin{algorithm*}[]
\DontPrintSemicolon
  Randomly initialize critic network $Q_{j}\left(o_{j}, a_{j} \mid \vartheta_{j}^{c}\right)$ and actor network $u_{j}\left(o_{j} \mid \vartheta_{j}^{p}\right)$ with NN weights $\vartheta_{j}^{c}$ and $\vartheta_{j}^{p}$ for privacy-sensitive, and critic network $Q_{l}\left(\boldsymbol{O}_{l} ; a_{l_{1}}, a_{l_{2}}, \cdots, a_{l_{{N}}} \mid \vartheta_{l}^{c}\right)$ and actor network $u_{l}\left(o_{l} \mid \vartheta_{l}^{p}\right)$ with NN weights $\vartheta_{l}^{c}$ and $\vartheta_{l}^{p} $for privacy-insensitive.\;
  Initialize global critic network parameter $\Theta$ based on the FedWgt strategy of formula (43)\;
  Initialize target critic network $\widetilde{Q}_{j}$ and actor NN $\widetilde{u}_{j}$ for privacy-sensitive, and $\bar{Q}_{l}$ and $\bar{u}_{l}$ for privacy-insensitive\;
  Initialize the reply buffer $\mathbb{D}$\;
     \For{episode $\epsilon=1$ to $\mathcal{K}$}    
        { 
            Initialize the action exploration noise $\mathcal{N}_{0}$ according to the noise function $f\left(x\right)$ subject to the conditions of formulas (31) and (32)\;
            Receive initial state $o_{j}$ and $\boldsymbol{O}_{l}$ for privacy-sensitive and privacy-insensitive, respectively\;
             \For{$t=1$ to $max-episode-length$}
                {
                    Select action $a_{k}=\left\{\begin{array}{l}
                    u_{j}\left(o_{j} \mid \vartheta_{j}^{p}\right)+\mathcal{N}_{0} \quad if \quad  \delta_{k}=1 \\
                    u_{l}\left(o_{l} \mid \vartheta_{l}^{p}\right)+\mathcal{N}_{0} \qquad otherwise \end{array}\right.$ w.r.t the current policy and exploration for each agent\;
                    Execute actions $a=\left(a_{1}, a_{2}, \cdots, a_{{M}+{N}}\right)$ ,observe reward $r$ and new state $o^{\prime}$ \;
                    Store tuple $\left(o,a,r,o^{\prime}\right)$ of each agent observation in reply buffer $\mathbb{D}$\;
                    $o \leftarrow o^{\prime}$\;
                    \For{agent $k=1$ to ${M}$ and $\zeta_{k}=1$}
                        {
                            Sample a random mini-batch of $\mathbb{S}$ samples $\left(o_{j_{k}}^{i}, a_{j_{k}}^{i}, r_{j_{k}}^{i}, o_{j_{k}}^{i+1}\right)$ from $\mathbb{D}$\;
                            The target value of equation (36) can be adjusted as:
                           $y_{j_{k}}^{i}=r_{j_{k}}^{i}+\gamma \tilde{Q}_{j_{k}}^{i+1}\left(o_{j_{k}}^{i+1}, \tilde{u}_{j_{k}}^{i+1}\left(o_{j_{k}}^{i+1} \mid \tilde{\vartheta}_{j_{k}}^{p}\right) \mid \tilde{\Theta}_{j_{k}}\right)$\;
                            Update critic by minimizing the loss $\mathcal{L}\left(\Theta_{j_{k}}\right)=\frac{1}{\mathbb{S}} \sum_{i}\left(y_{j_{k}}^{i}-Q_{j_{k}}^{i}\left(o_{j_{k}}^{i}, a_{j_{k}}^{i} \mid \Theta_{j_{k}}\right)\right)^{2}$\;
                            Update actor using sampled policy gradient:
                           $\nabla_{\vartheta_{j_{k}}^{p}} \mathcal{J} \approx \frac{1}{\mathbb{S}} \sum_{i} \nabla_{a} Q_{j_{k}}^{i}\left(o_{j_{k}}^{i}, a_{j_{k}}^{i} \mid \Theta_{j_{k}}\right)\mid_{o_{j_{k}}^{i} a_{j_{k}}^{i}=\mu\left(o_{j_{k}}^{i}\right)} \nabla_{\vartheta_{j_{k}}^{p}} \mu\left(o_{j_{k}}^{i} \mid \vartheta_{j_{k}}^{p}\right)\mid_{o_{j_{k}}^{i}}$
                        }
                    \For{agent $k=1$ to ${N}$ and $\zeta_{k}=0$}
                        {
                            Sample a random minibatch of $\mathbb{S}$ samples $\left(\boldsymbol{O}_{l_{k}}^{i}, \boldsymbol{a}_{l_{k}}^{i}, r_{l_{k}}^{i}, \boldsymbol{O}_{l_{k}}^{i+1}\right)$ from $\mathbb{D}$\;
                            The target value of equation (40) can be adjusted as:
                            $y_{l_{k}}^{i}=r_{l_{k}}^{i}+\gamma \bar{Q}_{l_{k}}^{i+1}\left(\boldsymbol{O}_{l_{k}}, \bar{a}_{l_{1}}, \bar{a}_{l_{2}}, \ldots, \bar{a}_{l_{{N}}} \mid \bar{\Theta}_{l_{k}}\right)\mid_{\bar{a}_{l_{c}}=\bar{u}_{l_{c}}\left(o_{j_{c}} \mid \bar{\vartheta}_{l_{c}}^{p}\right)}$\;
                            Update critic by minimizing the loss:
                            $\mathcal{L}\left(\Theta_{l_{k}}\right)=\frac{1}{\mathbb{S}} \sum_{i}\left(y_{l_{k}}^{i}-Q_{l_{k}}^{i}\left(\boldsymbol{O}_{l_{k}}^{i}, a_{l_{1}}, a_{l_{2}}, \cdots, a_{l_{{N}}} \mid \Theta_{l_{k}}\right)\mid_{a_{l_{c}}=u_{l_{c}}\left(o_{l_{c}} \mid \vartheta_{l_{c}}^{p}\right)}\right)^{2}$\;
                            Update actor using sampled policy gradient:
                           $\nabla_{\vartheta_{l_{k}}^{p}} \mathcal{K} \approx \frac{1}{S} \sum_{i} \nabla_{\vartheta_{l_{k}}^{p}} \mu_{l_{k}}\left(o_{l_{k}}^{i} \mid \vartheta_{l_{k}}^{p}\right) \nabla_{a_{l_{k}}} Q_{l_{k}}^{i}\left(\boldsymbol{O}_{l_{k}}^{i}, a_{1}^{i}, \ldots, a_{l_{k}}, \ldots, a_{l_{{N}}}^{i} \mid \Theta_{l_{k}}\right)\mid_{a_{l_{c}}=u_{l_{c}}\left(o_{l_{c}} \mid \vartheta_{l_{c}}^{p}\right)}$
                        }
                    Update the critic network parameter $\Theta$ based on the FedWgt strategy of formula (43)\;
                    Update target network parameters for each agent $i$ according formula (38)\;       
                }
        }
\caption{Joint optimization of wireless communication and computing resources allocation based on MAFWDRL}
\end{algorithm*} 

In order to address the disparity between distributed learning and centralized learning in wireless network optimization, the substitution of the currently learned model parameters $\vartheta_{j_{k}}$ and $\vartheta_{l_{k}}$ with $\Theta_{k}$ is contemplated for different learning architectures. By incorporating $\Theta_{k}$ into the training process, we aim to bridge the gap and compensate for the differences between the two learning paradigms.

The total complexity of the MAFWDRL algorithm is primarily determined by several key factors: the number of agents, the complexity involved in the NN's forward and backward propagation, the complexity associated with model integration, the interactions between the NN  and wireless systems, and the overall number of iterations. The complexity arising from the training process is influenced by the type of training mode employed. Specifically, for $M$ privacy-sensitive agents, the training complexity is given by $\mathrm{O}\left(4 M \times \sum_{p=1}^{P-1} B_{p} B_{p+1}\right)$, whereas for $N$ privacy-insensitive agents, it stands at $\mathrm{O}\left(2 (N+1) \times \sum_{p=1}^{P-1} B_{p} B_{p+1}\right)$. Here, $B_{p}$ denotes the number of neurons in the $p$-th layer of the network, and $ P$  represents the total number of layers in the NN. Furthermore, both the complexities of model integration and agent interactions exhibit a linear relationship with the number of iterations $\mathcal{K}$, represented as $O(\mathcal{K})$. Consequently, the overall complexity of the MAFWDRL algorithm can be accurately quantified as $\mathrm{O}\left((4 M+2 N+2) \mathcal{K} \sum_{p=1}^{P-1} B_{p} B_{p+1}+\mathcal{K}\right)$.

% \subsection{ }\label{AA}

\section{Simulation results }

\begin{figure}[]
  \begin{center}
  
  \subfigure[The 10th second]{
    \scalebox{0.17}[0.17]{\includegraphics{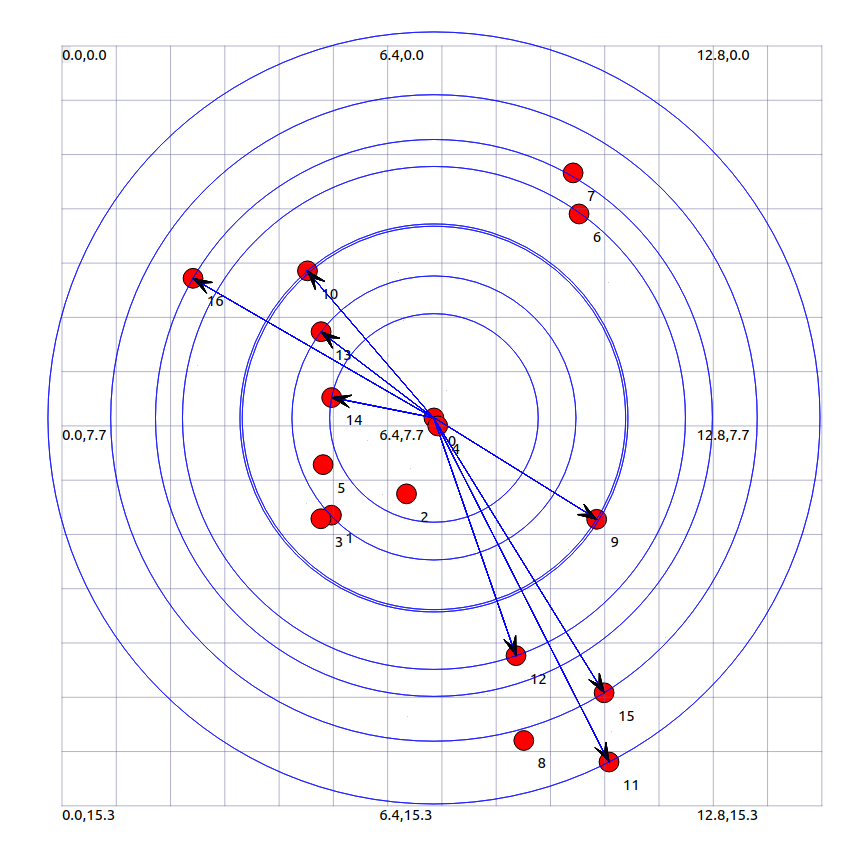}}
    
    \label{fig:2}
    }
     \subfigure[The 14th second]{
    \scalebox{0.17}[0.17]{\includegraphics{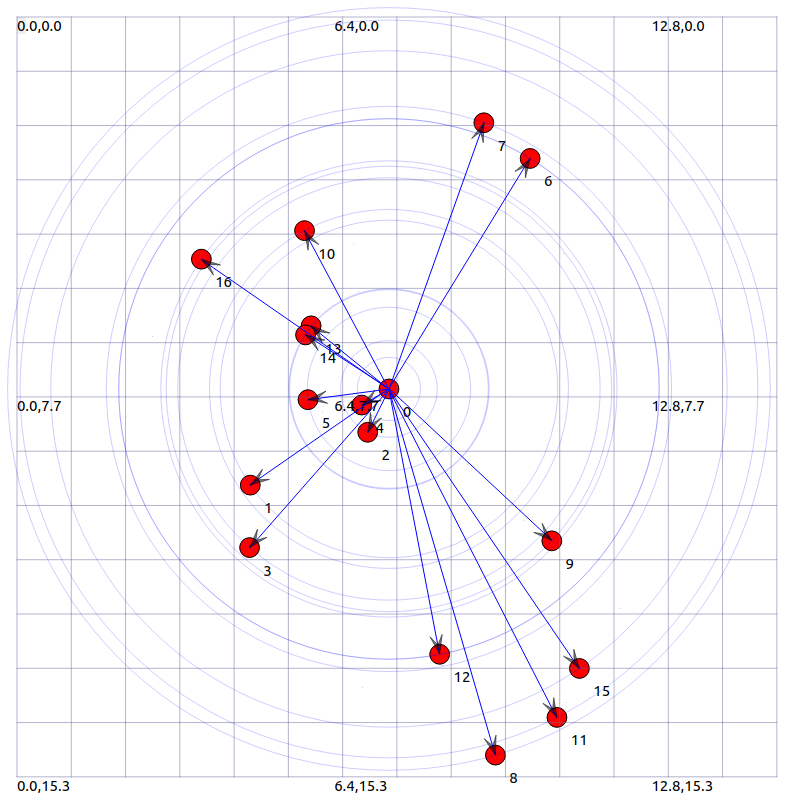}}
    \label{fig:3}
    }    \caption{An instance of user location and communication status simulated by NS3-simulator.}
  \end{center}
  
\end{figure}

The experiment involves the construction of a cross-platform collaboration simulation for NS3 (communication terminal) and Python (AI terminal) using NS3AI \cite{b14}. This simulation platform ensures the reality and reliability of the communication system simulation. The communication process between the STAs and the AP is fully captured by the NS3 terminal, spanning from the application layer to the physical layer. However, our focus lies in the adjustment of certain network parameters in the MAC layer through AI methods. Within this section, we present simulation results to validate the effectiveness of the proposed MAFWDRL algorithm for jointly optimizing computing and communication resource allocation in distributed wireless systems. The impact of adjusting specific parameters of the MAFWDRL algorithm and the execution strategy on wireless communication networks is considered in our analysis.

We construct a mobility model of wireless communication based on NS3-simulator to simulate UEs dynamic access control. We divide the task period $T$ into $L$ time slots to observe a variation of wireless channel state information within the time slot length $\tau =T⁄L$. Specifically, a wireless communication system is considered, comprising $1$ AP, $8$ privacy-sensitive STAs, and $8$ privacy-insensitive STAs. The AP is designed to support up to $4$ antennas, while each STA is equipped with $2$ antennas and is integrated with an individual AI model. Furthermore, the STAs are distributed randomly within a circular area of $7.5$ meters radius. Each STA is assigned a constant velocity of $v=1m/s$, with a maximum permissible movement radius of $20$ meters. Fig. 3 illustrates an instance of the visualization system built on the basis of the NS3-simulator. In Fig. 3 (a) and 3(b), we observe the spatial distribution of user locations and their communication statuses at the 10th and 14th seconds, respectively. The blue arrow symbolizes the connection between the STA and the AP at that specific moment, indicating the establishment of a communication link. Additionally, to meet the constraints imposed by the wireless communication environment, the actor and critic NN models are designed with $3$ hidden layers, with one hidden layer containing $8$ neurons and uses ReLU activation. This configuration is chosen deliberately to avoid adverse effects on the wireless features \cite{b15} and the risk of gradient vanishing. For a detailed summary of the remaining system parameters, please refer to Table I \cite{b8_a,b11,b16,b17}

\begin{table}[h]
\centering
\caption{Simulation parameters}%
\begin{tabular}{c c c c }
\hline %
\textbf{Parameters for Communication System} & {\textbf{Values}} \\
\hline%
Communication Frequency& 5GHz\\
\hline
Maximum Channel Bandwidth	& 80 MHz\\
\hline
AP/STA TX power	& 23/20 dBm\\
\hline
Antenna TX/RX gain & 3 dB	\\
\hline
Maximum CPU frequency of AP 	& $2 \times 10^{9}$ cycle/sec	\\
\hline
Maximum CPU frequency of STA	& $500 \times 10^{6}$ cycle/sec	\\
\hline
Effective Switched Capacitance &  $10^{-26}$	\\
\hline
Maximum Number of AP/STA Antennas 	& 4/2	\\
\hline
CPU Calculation Cycle of 1-bit	& 330 cycle/bit\\
\hline
FLOPS Calculation Rate of 1-cycle	& 8 flops/cycle\\
\hline
Number of Privacy-sensitive STAs & 8\\
\hline
Number of Privacy-insensitive STAs & 8\\
\hline 
Interaction step & 200 ms\\
\hline %
\textbf{Parameters for MAFDRL} & {\textbf{Values}} \\
\hline%
Memory Capacity & 100\\
\hline
Actor NN Learning Rate & 0.002\\
\hline
Critic NN Learning Rate & 0.02\\
\hline
Soft Update Parameter & 0.1\\
\hline
Mini-batch Size & 8\\
\hline
Discount rate & 0.1\\
\hline

\end{tabular}
\end{table}

\subsection{Experiments on noise design for off-policy MADRL}\label{AA}

In this section, our primary objective is to conduct a comprehensive and equitable assessment of noise adjustment's impact on the performance of distributed wireless systems. To achieve this, we have taken several measures. First, we have categorized all experimental users within the communication scenario as privacy non-sensitive, ensuring a level playing field for our evaluations. Additionally, we have employed the off-policy DRL algorithm MADDPG for our testing procedures.

To facilitate meaningful comparisons, we have considered two distinct noise adjustment functions. The first is the conventional noise linear decrease function, denoted as $f(t)=-\phi t+n_{0}$. The second is an illustrative example, which adheres to the criteria outlined in Section III, Subsection B, expressed as $g(t)=-(\eta t)^{3}+n_{0}$. Here, it is important to note that $n_{0}$ serves as a controlling factor, determining the maximum offset scale for actions that MADDPG can select. Meanwhile, $\phi$ and $\eta$ play crucial roles in regulating the magnitude of decrease in each iteration for the linear function $f(t)$ and the concave function $g(t)$, respectively. Our criteria for assessing system performance depend on the observed throughput of the entire system, which provides an objective basis for evaluating the various noise adjustment approaches.

Fig. 4 shows the performance comparison results of MADDPG in optimizing wireless communication throughput using the noise design method and the baseline scheme outlined in this paper. Specifically, Figs. 4(a) and 4(b) examine the impact of noise change functions, denoted as $f(t)$ and $g(t)$, at different offset scales ($n_{0}$) and various change rates ($\phi$ and $\eta$) on the performance of wireless communication networks. When comparing Figs. 4(a) and 4(b), it becomes apparent that the model exhibits greater stability when $n_{0}=1$ in comparison to when  $n_{0}=0.5$. This increased stability at  $n_{0}=1$ is attributed to its ability to provide a better foundation for the off-policy policy model and a broader exploratory space. The change rate, determined by $\phi$ and $\eta$, influences the distribution range of communication samples within the exploratory space. Optimal values for these change rates result in a richer set of samples in the memory replay.

However, it is crucial to strike a balance with the change rates. If $\phi$ and $\eta$ are set too low, the offset remains consistently at the maximum state, skewing the distribution of communication samples towards larger or smaller values. Conversely, excessively high change rates can limit exploration capabilities of the model, hindering effective training of communication models. For wireless communications, selecting $g(t)$ and setting the change rate $\eta=0.02$ leads to relatively stable model performance. Additionally, upon closer examination of Fig. 4, we observe that the state distribution of $g(t)$ during the initial exploration phase is broader compared to that of $f(t)$.

\begin{figure}[]
  \begin{center}
  \subfigure[Throughput against time step for offset scales $n_{0}=0.5$]{
    \scalebox{0.5}[0.45]{\includegraphics{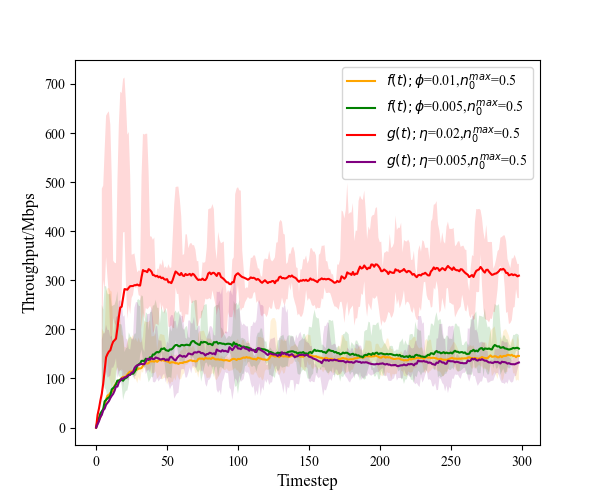}}
    
    \label{fig:2}
    }
     \subfigure[Throughput against time step for offset scales $n_{0}=1$]{
    \scalebox{0.5}[0.45]{\includegraphics{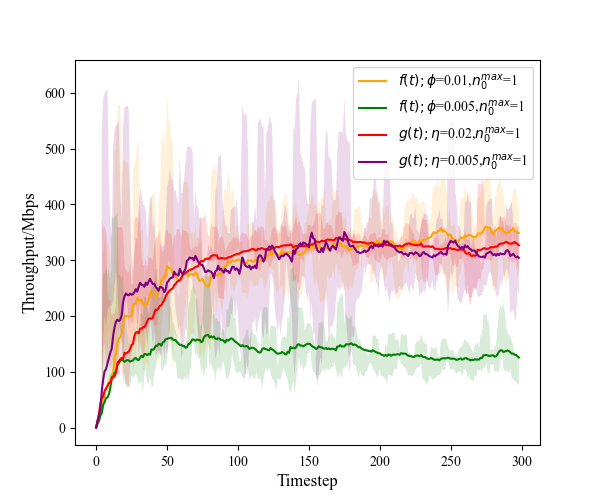}}
    \label{fig:3}
    }
    \caption{Performance comparison of the explore noise between baseline function $f$ and the proposed function $g$.}
  \end{center}
\end{figure}

\subsection{Experiments on MAFWDRL}\label{AA}

In this section, we conduct experiments involving scenarios depicted in Fig. 2, encompassing both privacy-sensitive and non-sensitive STAs. Utilizing $g(t)$ noise design function detailed in this paper, we conducted a comparative analysis of the application between the classic federated averaging (FedAvg) strategy for MADRL ($i.e.$, MAFADRL) and our proposed FedWgt strategy for MADRL ($i.e.$, MAFWDRL) under our proposed model architecture. Among them, these experiments were conducted under varying change rates of exploration function $g(t)$. Our main objective was to evaluate the overall system throughput as the reward feedback for our model. This measure provides a more intuitive reflection of the model's capability to optimize the communication network, and the results are presented in Fig. 5. We set the change rates $\eta$ to three different values, specifically $0.005$, $0.01$, and $0.02$, for the comparative analysis.

Our objectives were twofold: firstly, to validate the effectiveness of the $g(t)$ design and secondly, to evaluate the optimization performance of MAFWDRL in a distributed communication system. The results clearly indicate that our proposed method has substantially enhanced the overall system throughput while ensuring greater model stability when compared to MAFADRL. The heterogeneity causes variability in the performance of DRL models, a phenomenon that our experimental results confirm. Models employing the federated averaging strategy remain relatively stable, unaffected by $g(t)$. Conversely, our MAFWDRL method, as proposed in this paper, exhibits results consistent with those presented in Fig. 4, achieving superior model performance, particularly at $\eta= 0.02$.

\begin{figure}[]
  \begin{center}
    \scalebox{0.42}[0.4]{\includegraphics{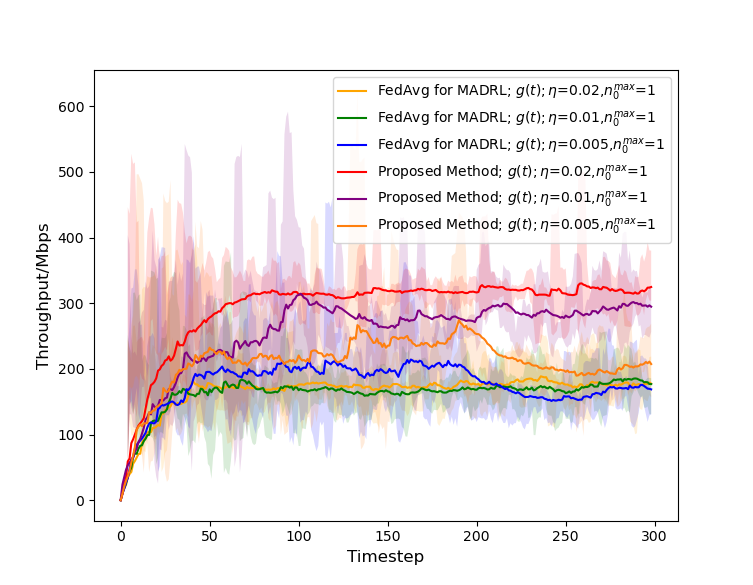}}
    \label{fig:3}
    \caption{Performance comparison of MAFADRL and the proposed MAFWDRL in term of the proposed noise function $g$ with different change rate $\eta$.}
  \end{center}
\end{figure}

% The common consensus suggests that the federated average strategy in the IID state yields better model performance. However, when considering the communication resources within wireless networks, we contend that there exists heterogeneity among the channel states of different STAs.

Fig. 6 presents the convergence results of the MAFWDRL model, which involves joint optimization of computing consumption and throughput. As illustrated in Fig. 6, several notable trends emerge. Initially, as the model iterates, the average loss of the agents shows a diminishing trend while fluctuating. Simultaneously, the average reward is increasing while fluctuating. After approximately 100 iterations, the model reaches a state of convergence. These trends observed during model training align closely with our ideal expectations. 

To thoroughly assess the optimization capabilities of MAFWDRL within distributed wireless networks and its adaptability to various real-world challenges, we conducted experiments employing two distinct reward settings:

\emph{1) Joint Optimization of Throughput Performance and Delay Priority:} Our first reward setting focuses on achieving the dual objectives of minimizing computing delay while maximizing throughput, defined as $r_{k}=\left\{\begin{array}{l}
z(\frac{\mathcal{T}_{k}^{\delta_{k}=0}}{{T}_{k}^{\delta_{k}=0}}) \\
z(\frac{\mathcal{T}_{k}^{\delta_{k}=1}}{{T}_{k}^{\delta_{k}=1}})
\end{array}\right.$. Essentially, we aim to strike a balance between minimizing the delay and maximizing system throughput. The objective is to adjust the reward to minimize computing delay and maximize throughput. 

\emph{2) Joint Optimization of Throughput and Computing Energy under Maximum Computing Delay Constraints:} Our second reward setting emphasizes the optimization of throughput and computing energy while adhering to a stringent constraint on the maximum permissible computing delay, as shown in Formula (34). In this case, we aim to jointly optimize throughput and computing energy while ensuring that the computed solutions adhere to the limitations imposed by Formula (34).

% \subsection{Performance Analysis of the Proposed MAFDRL Algorithm}\label{AA}

\begin{figure}[]
  \begin{center}
    \scalebox{0.55}[0.5]{\includegraphics{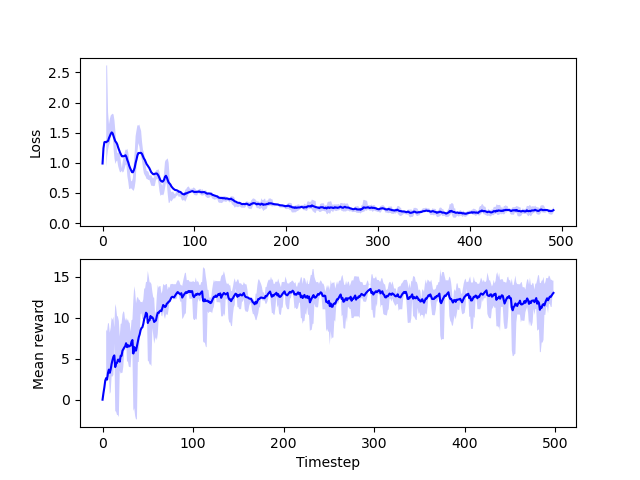}}
    \label{fig:3}
    \caption{Convergence performance of the proposed MAFWDRL optimization algorithm for throughput and energy consumption.}
  \end{center}
\end{figure}

Figs. 7, 8, and 9 depict the performance variations in model system throughput, calculation latency, and energy consumption of the MAFADRL and the proposed MAFWDRL under the aforementioned two reward settings. The maximum calculation delay constraint of reward 2) is set to $T_{max}=500 ms$. Figs. 7(a) and 8(a) illustrate the training curves for throughput and calculation delay under reward 1), while Figs. 7(b) and 8(b) display the training outcomes for reward 2).

In the performance comparison between Fig. 7(a) and Fig. 7(b), we observe that the convergence throughput of the system with reward 1) is approximately 500 Mbps, surpassing the 290 Mbps achieved with reward 2). The setting of this penalty mechanism restricts the throughput of the system to a certain extent to ensure that the system can meet the computing delay constraints. As for the reward scheme 1) (which is no computing delay constraint),  the agent will be guided to act in the direction of maximizing $\mathcal{O}^{’}=\frac{\mathcal{T}}{T}$. Therefore, reward 1) is more conducive to throughput improvement than reward 2). The simulation analysis results align with our theoretical analysis, thereby enabling the MAFWDRL model to effectively address the diverse communication needs for the future. Furthermore, in Fig. 7, we note that the throughput for both privacy-sensitive and privacy-insensitive users remains relatively consistent. This consistency arises from the federated weighted policy, which synchronously updates all training model parameters on the server and transmits them to the agent within the same communication environment. This approach allows the model to compensate for differences between privacy-sensitive and privacy-insensitive training modes. From the perspective of algorithm application, the simulation results show that the throughput performance of the proposed FedWgt strategy is significantly better than that of the FedAvg strategy in scheme 1) and scheme 2). Theoretical analysis shows that the greater the difference between agents, the better the FedWgt strategy is than the FedAvg strategy, and the model has stronger generalization ability.

% System throughput is positively related to task calculation delay, so the reward scheme 2) makes the agent give priority to meeting the delay constraint requirements before optimizing the target $\mathcal{O}=\frac{\mathcal{T}}{E}$.

\begin{figure}[]
  \begin{center}
  \subfigure[Reward scheme 1]{
    \scalebox{0.38}[0.33]{\includegraphics{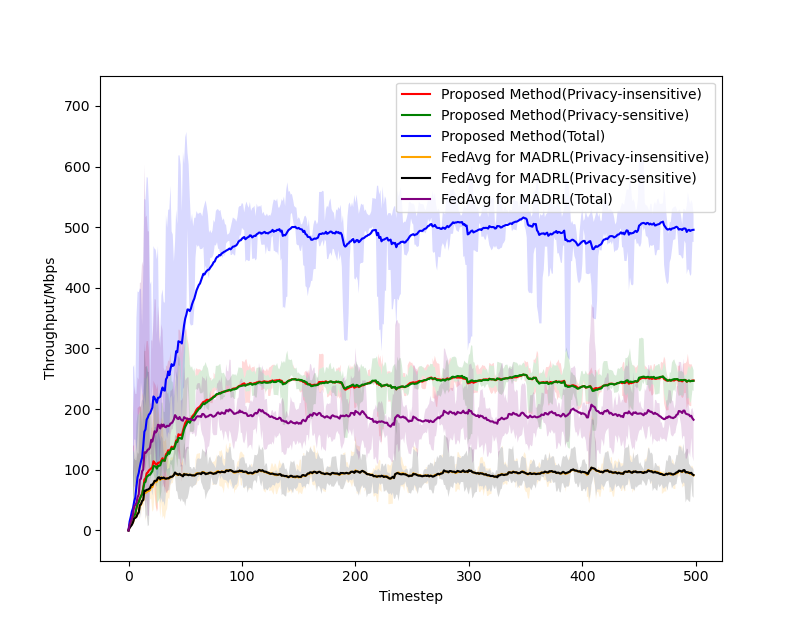}}
    \label{fig:2}
    }
    
  \subfigure[Reward scheme 2]{
    \scalebox{0.40}[0.33]{\includegraphics{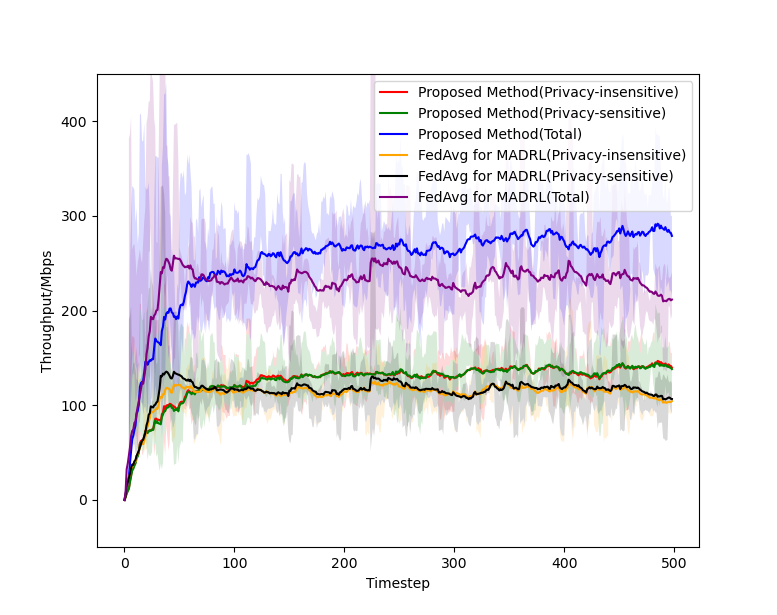}}
    \label{fig:3}
    }
    \caption{Throughput performance comparison of the MAFADRL and the proposed MAFWDRL optimization algorithm for different reward design scheme.}
  \end{center}
\end{figure}

\begin{figure*}[]
  \begin{center}
  \subfigure[Average latency of all STAs based on reward scheme 1]{
    \scalebox{0.38}[0.39]{\includegraphics{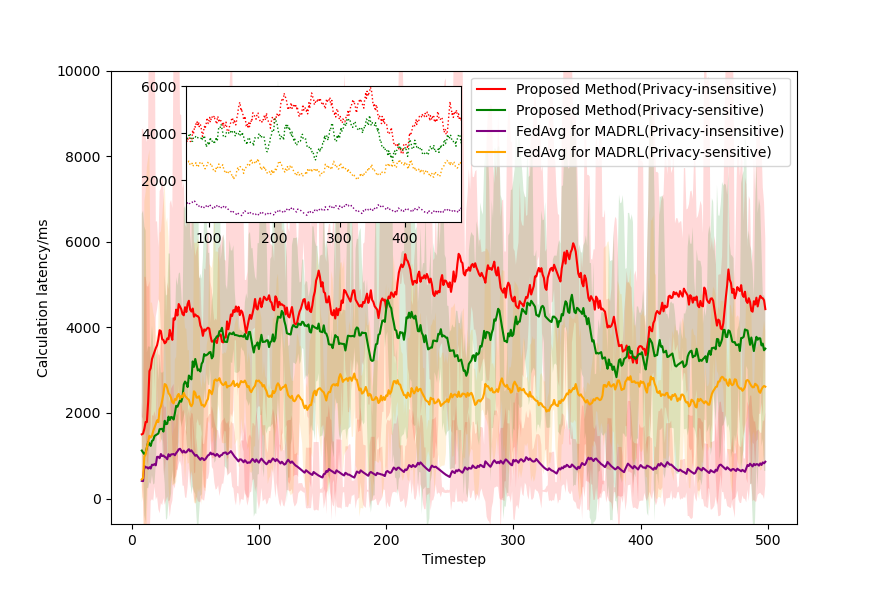}}  
    \label{fig:2}
    }
    \subfigure[STA convergence latency based on reward scheme 1]{
    \scalebox{0.43}[0.43]{\includegraphics{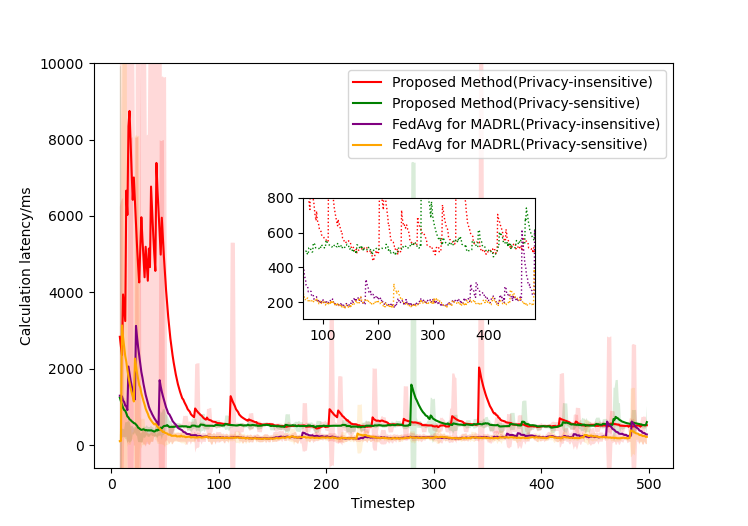}}  
    \label{fig:2}
    }
    \subfigure[Average latency of all STAs based on reward scheme 2]{
    \scalebox{0.42}[0.35]{\includegraphics{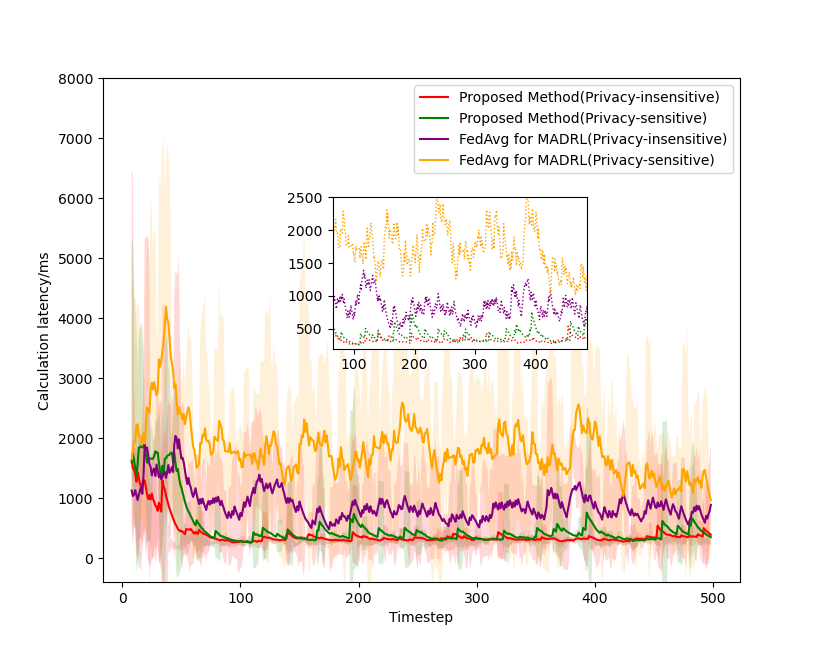}}
    \label{fig:3}
    }
    \subfigure[STA convergence latency based on reward scheme 2]{
    \scalebox{0.42}[0.48]{\includegraphics{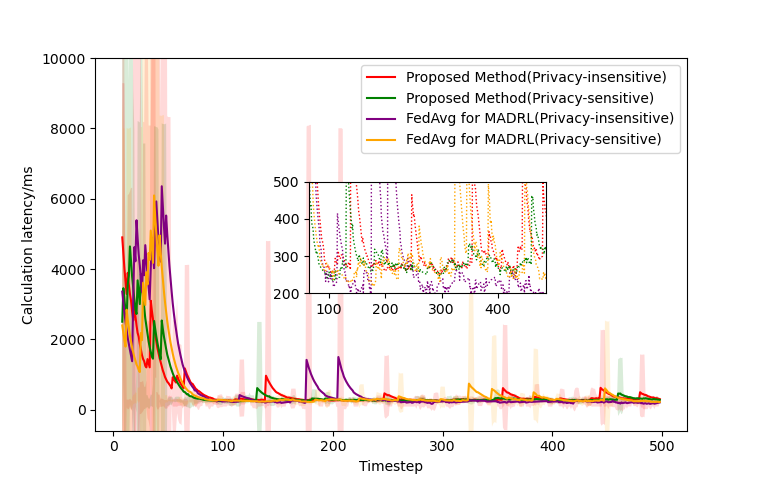}}  
    \label{fig:2}
    }
    \caption{Calculation latency performance comparison of the MAFADRL and the proposed MAFWDRL optimization algorithm for different reward design schemes.}
  \end{center}
\end{figure*}

% Since computing delay is directly linked to throughput, the reward setting prioritizing delay competes with throughput. This lower throughput can yield positive benefits in terms of latency, thereby constraining throughput performance. Reward 2) is adjusted to prioritize a system-focused energy reward strategy once the delay constraint is satisfied. Our analysis reveals that energy ($E$) is proportionate to both data quantity and the square of computational frequency ($f$). Consequently, the impact of $f$ is more pronounced than that of data volume, causing the computational delay of the system to approach $T_{max}$ from the left side of the X-axis. In this scenario, the increase in throughput associated with reward 2) leads to more favorable system growth compared to reward 1). 

\begin{figure}[]
  \begin{center}
    \scalebox{0.5}[0.45]{\includegraphics{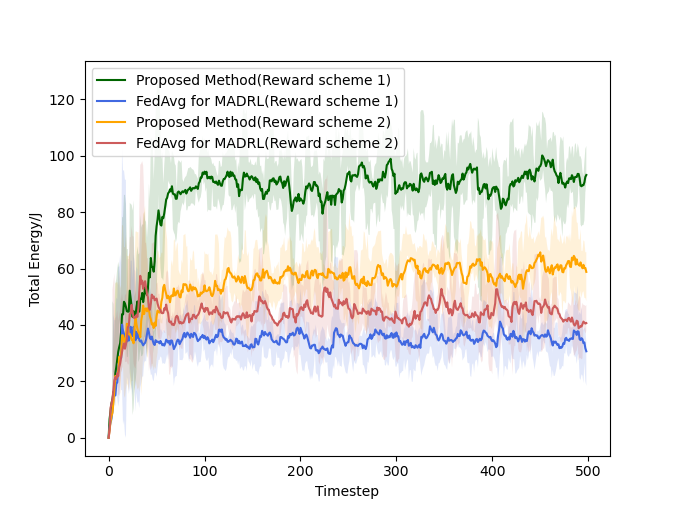}}
    \label{fig:3}
    \caption{Energy consumption comparison of the MAFADRL and the proposed MAFWDRL optimization algorithm for different reward design schemes.}
  \end{center}
\end{figure}

% Fig. 8 shows the simulation results of rewards 1) and 2) in calculating the delay change. Both privacy-sensitive and privacy-insensitive STAs can converge to a lower latency, which achieves the ideal situation. The computing delays for privacy-insensitive STAs of Fig.8 (a) and 8(b) are stable at about 285ms and 310ms respectively, and the privacy-sensitive STAs are stable at 300ms and 320ms which result is consistent with the previous analysis. After satisfying the delay constraint, the improvement of reward 2) delay will help improve the throughput, so the overall training delay of reward 1) is lower than reward 2).

% Fig. 8 shows the simulation results of the calculation latency of the agent changes based on reward schemes 1) and 2). Overall, both privacy-sensitive STAs and privacy-insensitive STAs can converge to lower delays. Specifically, for privacy-insensitive STAs, the calculation latencies stabilize at approximately 550 ms and 300 ms in Figs. 8(a) and 8(b), respectively. Meanwhile, the latency for privacy-sensitive STAs consistently remains around 600 ms and 320 ms. Although reward scheme 1) effectively maintains a relatively low latency level and enhances system throughput, it falls short in ensuring latency stability across all agents. Conversely, reward scheme 2) demonstrates greater stability in calculation latency, achieving lower latency. Notably, the computing latency for both privacy-sensitive and privacy-insensitive users adheres to the established constraint requirement of $T<T_{max}=500ms$. 

Fig. 8 present the simulation results for the calculation latency of both MAFADRL and our proposed MAFWDRL models, utilizing two distinct reward schemes. Figs 8(a) and 8(c) depict the global average latency and convergence latency achieved under reward scheme 1, while Figs. 8(b) and 8(d) illustrate the outcomes for reward scheme 2. The global average latency is indicative of the multi-agent algorithm's stability, whereas convergence latency reflects the optimal performance attainable by the model. Under reward scheme 2, the MAFWDRL exhibits superior performance in terms of global average latency, maintaining stability within a 500 ms latency constraint, thereby demonstrating enhanced algorithmic stability. Conversely, the MAFADRL achieves a lower convergence latency, correlated to its reduced throughput compared to MAFWDRL. With reward scheme 1, both MAFWDRL and MAFADRL display not stable enough for the wireless system.  MAFWDRL's higher system throughput results in a larger optimal convergence latency which absent a latency constraint. Furthermore, our simulations reveal that under the architecture designed in this study, both FedAvg and FedWgt enable agents to converge to an improved system state. 

% Fig.9 shows the training results of the total energy consumption of the MAFWDRL algorithm under different reward schemes. The energy consumption is mainly determined by f and the amount of data. Since the difference between reward 1) and reward 2) is small in the adjustment of $f$, the total data amount of reward 2) is larger than that of reward 1), which makes the energy consumption slightly higher. The results show that the energy fluctuation is maintained at a suitable level, achieving a more ideal effect.

% Fig.9 shows the training results of the total energy consumption of the MAFWDRL algorithm under different reward schemes. The energy consumption is mainly determined by CPU frequency $f$ and the amount of data. The difference between reward 1) and reward 2) is small in the adjustment of $f$, but the total data amount of reward 1) is larger than that of reward 2), which makes the energy consumption higher. The simulation results show that the energy fluctuation of reward 2) is maintained at a suitable level, achieving a more ideal effect. However, reward 1) has better throughput than reward 2). 

 Fig.s 7 and 8 present the results that form the basis for the analysis in Fig. 9, which illustrates the total energy consumption of both MAFADRL and the proposed MAFWDRL. It is observed that MAFWDRL demonstrates superior system throughput compared to MAFADRL. Consequently, the lower bound for the total energy consumption of MAFWDRL is expected to be higher than that of MAFADRL. Moreover, the trend in total energy consumption depicted in Fig. 9 is consistent with the patterns observed in Fig. 7, further corroborating this conclusion.

\section{Conclusion}

In this paper, we present a novel wireless intelligent distributed communication system based on MAFWDRL. The system comprises two core modules: the MADRL module and the FedWgt module. To address the crucial issue of privacy security, we have devised distinct MADRL training modes for various privacy categories. These training modes effectively tackle the challenge of distributed wireless resource allocation while ensuring robust user privacy protection. However, one limitation of this training approach is that it mainly concentrates on the local communication status of individual agents, overlooking the intricate interplay of communication resource allocation across the entire distributed system. To overcome this limitation, we introduce a FedWgt strategy that considers the diverse training needs of each agent model while protecting the security of privacy. This innovative approach optimizes the global critic model, endowing it with the versatility to accommodate different privacy types. Additionally, we delve into the design of the noise curve within the off-policy DRL model, enabling the model to explore more extensively, enriching the experiment pool of the system, and elevating model optimization performance. The results of our simulation experiments underscore the substantial enhancements achieved by our MAFWDRL model. Notably, it significantly boosts the throughput of the entire communication system, and diminishes computing latency and energy consumption, all while maintaining stringent privacy security requisites.

Nevertheless, the intelligent optimization of wireless communication scenarios continues to confront a range of increasingly complex practical challenges. Looking forward, our research will delve into both intention-driven and data-driven generative AI-aided approaches within wireless communication systems \cite{niyato2,niyato3 }, which aims to enhance the model's generalization capabilities, adaptability, and interpretability. These research are poised to expedite the evolution of intelligent communications, aligning them with the demands of the forthcoming 6G intelligent era.

\bibliographystyle{IEEEtran}
\bibliography{reference}

\vspace{12pt}
% \color{red}
% IEEE conference templates contain guidance text for composing and formatting conference papers. Please ensure that all template text is removed from your conference paper prior to submission to the conference. Failure to remove the template text from your paper may result in your paper not being published.

\end{document}